\documentclass[authoryear,11pt]{elsarticle}%
\usepackage[flushleft]{threeparttable}
\usepackage{ragged2e}
\usepackage{lscape}
\usepackage{amsmath}
\usepackage{amsthm}
\usepackage{graphicx}
\usepackage{amsfonts}
\usepackage{amssymb}
\usepackage{scalefnt}

\usepackage{natbib}
\usepackage{hyperref}

\usepackage{setspace}
\usepackage{array}%
\providecommand{\U}[1]{\protect\rule{.1in}{.1in}}
\numberwithin{equation}{section}
\theoremstyle{plain}
\newtheorem{theorem}{Theorem}
\newtheorem{lemma}{Lemma}
\newtheorem{assumption}{Assumption}

\theoremstyle{remark} 
\newtheorem{definition}{Definition}

\advance \topmargin by -\headheight
\advance \topmargin by -\headsep
\evensidemargin \oddsidemargin
\newcolumntype{L}[1]{>{\raggedright\let\newline\\arraybackslash\hspace{0pt}}m{#1}}
\newcolumntype{C}[1]{>{\centering\let\newline\\arraybackslash\hspace{0pt}}m{#1}}
\newcolumntype{R}[1]{>{\raggedleft\let\newline\\arraybackslash\hspace{0pt}}m{#1}}

\begin{document}

\begin{frontmatter}
\title{Fixed-$b$ Asymptotics for Panel Models with Two-Way Clustering}
\author[inst1]{Kaicheng Chen}
\affiliation[inst1]{organization={Department of Economics, Michigan State University},
            addressline={East Lansing, MI 48824},
           country={United States of America}}
\author[inst1,*]{Timothy J. Vogelsang}
\affiliation[*]{Corresponding author. E-mail address: tjv@msu.edu}
\begin{abstract}
This paper studies a cluster robust variance estimator proposed by
Chiang, Hansen and Sasaki (2024) for linear panels. First, we show algebraically that this
variance estimator (CHS estimator, hereafter) is a linear combination of three
common variance estimators: the one-way unit cluster estimator, the
\textquotedblleft HAC of averages\textquotedblright\ estimator, and the
\textquotedblleft average of HACs\textquotedblright\ estimator. Based on this
finding, we obtain a fixed-$b$ asymptotic result for the CHS estimator and
corresponding test statistics as the cross-section and time sample sizes
jointly go to infinity. Furthermore, we propose two simple bias-corrected
versions of the variance estimator and derive the fixed-$b$ limits. In a
simulation study, we find that the two bias-corrected variance estimators
along with fixed-$b$ critical values provide improvements in finite sample
coverage probabilities. We illustrate the impact of bias-correction and use of
the fixed-$b$ critical values on inference in an empirical example on the relationship between industry profitability and market concentration.

\end{abstract}
\end{frontmatter}

\setcounter{page}{1} \thispagestyle{empty} \pagestyle{plain}

\section{Introduction}

\label{sec:introduction}When carrying out inference in a linear panel model,
it is well known that failing to adjust the variance estimator of estimated
parameters to allow for different dependence structures in the data can cause
over-rejection/under-rejection problems under null hypotheses, which in turn
can give misleading empirical findings (see \citealp{Bertrand2004}).

To study different dependence structures and robust variance estimators in
panel settings, it is now common to use a component structure model
$y_{it}=f(\alpha_{i},\gamma_{t},\varepsilon_{it})$ where the observable data,
$y_{it}$, is a function of an individual component, $\alpha_{i}$, a time
component, $\gamma_{t}$, and an idiosyncratic component, $\varepsilon_{it}$.
See, for example, \cite{Davezies2021},
\cite{mackinnon2021wild}, \cite{Menzel2021}, and \cite{CHS_Restat}. As a
concrete example, suppose $y_{it}=\alpha_{i}+\varepsilon_{it}$ for $i=1,...,N$
and $t=1,...,T$, where $\alpha_{i}$ and $\varepsilon_{it}$ are assumed to be
i.i.d random variables. The existence of $\alpha_{i}$ generates serial
correlation within group $i$, which is also known as the individual clustering
effect. This dependence structure is well-captured by the cluster variance
estimator proposed by \cite{Liang1986} and \cite{Arellano1987}. One can also
use the \textquotedblleft average of HACs" variance estimator that uses
cross-section averages of the heterogeneity and autocorrelation (HAC) robust
variance estimator proposed by \cite{Newey1987}. On the other hand, suppose
$y_{it}=\gamma_{t}+\varepsilon_{it}$ where $\gamma_{t}$ is assumed to be an
i.i.d sequence of random variables. Cross-sectional/spatial dependence is
generated in $y_{it}$ by $\gamma_{t}$ is through the time clustering effect.
In this case one can use a variance estimator that clusters over time or use
the spatial dependence robust variance estimator proposed by
\cite{Driscoll1998}. Furthermore, if both $\alpha_{i}$ and $\gamma_{t}$ are
assumed to be present, e.g. $y_{it}=\alpha_{i}+\gamma_{t}+\varepsilon_{it}$,
then the dependence of $\{y_{it}\}$ exists in both the time and cross-section
dimensions, also as known as two-way clustering effects. Correspondingly, the
two-way/multi-way robust variance estimator proposed by
\cite{ColinCameron2011} is suitable for this case.

In macroeconomics, the time effects, $\gamma_{t}$, can be regarded as common
shocks which are usually serially correlated. Allowing persistence in
$\gamma_{t}$ up to a known lag structure, \cite{Thompson2011} proposed a
truncated variance estimator that is robust to dependence in both the
cross-section and time dimensions. Because of unsatisfying finite sample
performance of this rectangular-truncated estimator, \cite{CHS_Restat} propose
a Bartlett kernel variant (CHS variance estimator, hereafter) and establish
validity of tests based on this variance estimator using asymptotics with the
cross-section sample size, $N$, and the time sample size, $T$, jointly going
to infinity. The asymptotic results of the CHS variance estimator rely on the
assumption that the bandwidth, $M$, goes to infinity as $T$ goes to infinity
while the bandwidth sample size ratio, $b=\frac{M}{T}$ is of a small order. As
pointed out by \cite{neave1970improved} and \cite{Kiefer2005}, the value of
$b$ in a given application is a non-zero number that matters for the sampling
distribution of the variance estimator. Treating $b$ as shrinking to zero in
the asymptotics may miss some important features of finite sample behavior of
the variance estimator and test statistics. As noted by \cite{Andrews1991},
\cite{Kiefer2005}, and many others, HAC robust tests tend to over-reject in
finite samples when standard critical values are used. This is especially true
when time dependence is persistent and large bandwidths are used. We document
similar findings for tests based on CHS variance estimator in our simulations.

To improve the performance of tests based on CHS variance estimator, we derive
fixed-$b$ asymptotic results (see \citealp{Kiefer2005}, \citealp{sun2008optimal},
\citealp{Vogelsang2012}, \citealp{zhang2013fixed}, \citealp{sun2014let},
\citealp{bester2016fixed}, and \citealp{lls2021}). Fixed-$b$ asymptotics captures some
important effects of the bandwidth and kernel choices on the finite sample
behavior of the variance estimator and tests and provides reference
distributions that can be used to obtain critical values that depend on the
bandwidth (and kernel). Our asymptotic results are obtained for $N$ and $T$
jointly going to infinity and leverage the joint asymptotic framework
developed by \cite{Phillips1999}. The limiting distribution of tests based on
the CHS or BCCHS estimator are not asymptotically pivotal, so we propose a
plug-in method of simulating fixed-$b$ critical values. One key finding is
that the CHS variance has a multiplicative bias given by $1-b+\frac{1}{3}%
b^{2}\leq1$ resulting in a downward bias that becomes more pronounced as the
bandwidth increases. By simply dividing the CHS variance estimator by
$1-b+\frac{1}{3}b^{2}$ we obtain a simple bias-corrected variance estimator
that improves the performance of tests based on the CHS variance estimator
even without using plug-in fixed-$b$ critical values. We label this
bias-corrected CHS variance estimator as BCCHS.

As a purely algebraic result, we show that the CHS variance estimator is the
sum of the Arellano cluster and Driscoll-Kraay variance estimators minus the
\textquotedblleft averages of HAC\textquotedblright\ variance estimator. We
show that dropping the \textquotedblleft averages of HAC\textquotedblright%
\ component in conjunction with bias correcting the Driscoll-Kraay component
removes the asymptotic bias in the CHS variance estimator and has the same
fixed-$b$ limit as the BCCHS variance estimator. We label the resulting
variance estimator of this second bias correction approach as the DKA
(Driscoll-Kraay+Arellano) variance estimator. Similar ideas are also used by
\cite{davezies2018asymptotic} and \cite{mackinnon2021wild} where they argue
the removal of the negative and small order component in the variance
estimator brings computational advantage in the sense that the variance
estimates are ensured to be positive semi-definite. In our simulations we find
that negative CHS variance estimates can occur up to 6.4\% of the time. An advantage of the OKA variance estimator is guaranteed positive semi-definiteness. The DKA variance estimator also tends to deliver tests with
better finite sample coverage probabilities although there are exceptions:
when the data is independent and identically distributed (i.i.d.) in both the
cross-section and time dimensions, we show the DKA estimator has a different
fixed-$b$ limit and results in tests that are conservative including the case
where the bandwidth is small\footnote{In the small bandwidth case we find that
the limit of the DKA variance estimator is twice as big as the population
variance\ for i.i.d.\ data - a finding similar to Theorem 2 in
\cite{mackinnon2021wild} in a multiway clustering setting. We thank a referee
for pointing out the similarity between our results for the DKA variance
estimator and the results in \cite{mackinnon2021wild} for multiway cluster
variance estimators when the data is i.i.d.}. The fixed-$b$ limit of the CHS
variance estimator is also different in the i.i.d.\ case but tests based on it
remain robust when the bandwidth is small.

In a finite sample simulation study, we compare sample coverage probabilities
of confidence intervals based on CHS, BCCHS, and DKA variance estimators using
critical values from both the standard normal distribution and the fixed-$b$
limits. The fixed-$b$ limits of the test statistics constructed by these three
variance estimators are not pivotal, so we use a simulation method to obtain
the critical values via a plug-in estimator approach to handle asymptotic
nuisance parameters. While the plug-in fixed-$b$ critical values can
substantially improve coverage rates relative to using standard critical
values when using the CHS variance estimator, improvements from simply using
the bias corrections are impressive. In the case of data-dependent bandwidths,
the plug-in fixed-$b$ critical values provide further improvements in finite
sample coverage probabilities when neither $T$ nor $N$ is large. Conversely,
when both $N$ and $T$ are very small, bias correction alone can give more
accurate finite sample coverage probabilities than bias correction with
plug-in fixed-$b$ critical values. Similar results hold for tests based on the
DKA variance estimator.

Overall, four different approaches for within- and across-cluster dependent
robust tests are proposed: two are simple bias correction by BCCHS and DKA
estimators and the other two are bias correction by BCCHS and DKA estimators
with plug-in fixed-$b$ critical values. Even though tests based on BCCHS and
DKA are asymptotically equivalent under the main assumptions, their finite
sample performance is distinguishable. Based on theory and
simulation results, we provide comprehensive empirical guidance hinged on the
researcher's assessment of the data, model and priority of the test.

The rest of the paper is organized as follows. In Section 2 we sketch the
algebra of the CHS estimator and rewrite it as a linear combination of three
well-known variance estimators. In Section 3 we derive fixed-$b$ limiting
distributions of CHS-based tests for pooled ordinary least squares (POLS)
estimators in a simple location panel model and a linear panel regression
model. In Section 4 we derive the fixed-$b$ asymptotic bias of the CHS
estimator and propose two bias-corrected variance estimators. We also derive
fixed-$b$ limits for tests based on the bias-corrected variance estimators.
When the data is i.i.d.\ a key assumption for our asymptotic results no longer
holds and we show that the asymptotic limits change in this case. Section 5
presents finite sample simulation results that illustrate the relative
performance of $t$-tests based on the variance estimators. Some theoretical
results for two-way-fixed-effects (TWFE) estimator are also discussed along
with the simulation. In Section 6 we illustrate the practical implications of
the bias corrections and use of fixed-$b$ critical values in an empirical
example. Section 7 concludes the paper with guidance for empirical practice
and a discussion on the limitations of proposed approaches.

\section{A Variance Estimator Robust to Two-Way Clustering}

We first motivate the estimator of the asymptotic variance of the pooled
ordinary least squares (POLS) estimator under arbitrary dependence in both the
time and cross-section dimensions. Consider the linear panel model%
\begin{equation}
y_{it}=x_{it}^{\prime}\beta+u_{it},\ \ i=1,\dots,N,.\ \ t=1,\dots,T,
\label{reg}%
\end{equation}
where $y_{it}$ is the dependent variable, $x_{it}$ is a $k\times1$ vector of
covariates, $u_{it}$ is the error term, and $\beta$ is the coefficient vector.
Let $\widehat{\beta}$ be the POLS estimator of $\beta$. For illustrative
purposes the variance of $\widehat{\beta}$ can be approximated as%
\[
\text{Var}\left(  \widehat{\beta}\right)  \approx\widehat{Q}^{-1}\Omega
_{NT}\widehat{Q}^{-1},
\]
where $\widehat{Q}:= \frac{1}{NT}\sum_{i=1}^{N}\sum_{t=1}^{T}x_{it}%
x_{it}^{\prime}$ and $\Omega_{NT}:= \text{Var}\left(  \frac{1}{NT}\sum_{i=1}%
^{N}{\sum_{t=1}^{T}{v_{it}}}\right)  $ with $v_{it}:= x_{it}u_{it}$.

Without imposing assumptions on the dependence structure of $v_{it}$, it has
been shown, algebraically, that $\Omega_{NT}$ has the following form (see
\citealp{Thompson2011} and \citealp{CHS_Restat}):
\begin{align*}
\Omega_{NT}=  &  \frac{1}{N^{2}T^{2}}\left[  \sum\limits_{i=1}^{N}%
{\sum\limits_{t=1}^{T}{\sum\limits_{s=1}^{T}{\text{E}\left(  v_{it}%
v_{is}^{\prime}\right)  }}}+\sum\limits_{t=1}^{T}{\sum\limits_{i=1}^{N}%
{\sum\limits_{j=1}^{N}{\text{E}(v_{it}v_{jt}^{\prime}})}}-\sum\limits_{t=1}%
^{T}{\sum\limits_{i=1}^{N}{\text{E}\left(  v_{it}v_{it}^{\prime}\right)  }%
}\right. \\
&  +\sum\limits_{m=1}^{T-1}\sum\limits_{t=1}^{T-m}{\text{E}\left(
\sum\limits_{i=1}^{N}{v_{it}}\right)  \left(  \sum\limits_{j=1}^{N}%
{v_{j,t+m}^{\prime}}\right)  }-\sum\limits_{i=1}^{N}{\sum\limits_{t=1}%
^{T-m}{\text{E}\left(  v_{it}v_{i,t+m}^{\prime}\right)  }}\\
&  \left.  +\sum\limits_{m=1}^{T-1}\sum\limits_{t=1}^{T-m}{\text{E}\left(
\sum\limits_{i=1}^{N}{v_{i,t+m}}\right)  \left(  \sum\limits_{j=1}^{N}%
{v_{j,t}^{\prime}}\right)  }-\sum\limits_{i=1}^{N}{\sum\limits_{t=1}%
^{T-m}{\text{E}\left(  v_{i,t+m}v_{i,t}^{\prime}\right)  }}\right]  .
\end{align*}
Based on this decomposition of $\Omega_{NT}$, \cite{Thompson2011} and
\cite{CHS_Restat} each propose a truncation-type variance estimator. In
particular, \cite{CHS_Restat} replaces the \cite{Thompson2011} truncation
scheme with a Bartlett kernel and establish the consistency result of their variance estimator while allowing two-way clustering effects with serially correlated stationary time effects. 

As an asymptotic approximation, appealing to consistency of the estimated
variance allows the asymptotic variance to be treated as known when generating
asymptotic critical values for inference. While convenient, such a consistency
result does not capture the impact of the choice of $M$ and kernel function on
the finite sample behavior of the variance estimator and any resulting size
distortions of test statistics. To capture some of the finite sample impacts
of the choice of $M$ and kernel, we apply the fixed-$b$ approach of
\cite{Kiefer2005}.

Noticeably, the CHS variance estimator can be decomposed into three well-known variance estimators, which will be helpful when we apply the fixed-$b$ approximation. Using straightforward algebra, one can show that the CHS variance estimator defined in equation (2.12) of \cite{CHS_Restat} can be rewritten as
\begin{align}
    \widehat{V}_\text{CHS} := & \widehat{Q}^{-1} \widehat{\Omega}_\text{CHS} \widehat{Q}^{-1},  \label{var_chs} \\
    \widehat{\Omega}_\text{CHS}:= & \widehat{\Omega}_{A} + \widehat{\Omega}_\text{DK} -\widehat{\Omega}_\text{NW} , \label{omega_chs} 
\end{align}
where, with the Bartlett kernel defined as $k\left(  \frac{m}{M}\right)  =1-\frac{m}{M}$ and $M$ being the truncation parameter,
\begin{align}
{\widehat{\Omega}}_\text{A} := & \frac{1}{N^{2}T^{2}}\sum\limits_{i=1}^{N}%
{\sum\limits_{t=1}^{T}{\sum\limits_{s=1}^{T}{\left(  \hat{v}_{it}%
\hat{v}_{is}^{\prime}\right)  ,}}} \label{chs_a} \\
{\widehat{\Omega}}_\text{DK}:= & \frac{1}{N^{2}T^{2}}\sum\limits_{t=1}^{T}{\sum\limits_{s=1}^{T}{k\left(
\frac{\left\vert t-s\right\vert }{M}\right)  \left(  \sum\limits_{i=1}%
^{N}{{\hat{v}}_{it}}\right)  \left(  \sum\limits_{j=1}^{N}{{\hat{v}%
}_{js}^{\prime}}\right)  ,}}\label{chs_dk} \\
{\widehat{\Omega}}_\text{NW}:=  &  \frac{1}{N^{2}T^{2}}\sum\limits_{i=1}^{N}{\sum\limits_{t=1}^{T}{\sum\limits_{s=1}^{T}{k\left(  \frac{\left\vert t-s\right\vert }{M}\right)
{\hat{v}}_{it}{\hat{v}}_{is}^{\prime}.}}} \label{chs_nw}%
\end{align}
Notice that (\ref{chs_a}) is the \textquotedblleft cluster by
individuals\textquotedblright\ estimator proposed by \cite{Liang1986} and
\cite{Arellano1987}, (\ref{chs_dk}) is the \textquotedblleft HAC of cross-section
averages" estimator proposed by \cite{Driscoll1998}, and (\ref{chs_nw}) is the
\textquotedblleft average of HACs\textquotedblright\ estimator (see
\citealp{Petersen2009} and \citealp{Vogelsang2012}). In other words, ${\widehat{\Omega}}_\text{CHS}$ is a linear combination of three
well-known variance estimators that have been proposed to handle particular
forms of dependence structure. While there are some existing asymptotic
results for the components in (\ref{omega_chs}) that are potentially relevant (e.g.
\citealp{Hansen2007}, \citealp{Vogelsang2012}, and \citealp{CHS_Restat}), these results
are derived either under one-way dependence or are not sufficiently
comprehensive to directly obtain a fixed-$b$ result for ${\widehat{\Omega}%
}_\text{CHS}$. Some new theoretical results are needed.

\section{Fixed-$b$ Asymptotic Results}

\subsection{The Multivariate Mean Case}

\noindent To set ideas and intuition we first focus on a simple panel mean
model (panel location model) of a $k\times1$ random vector $y_{it}$ and then
extend the analysis to the linear regression case. We use a large-$N$ and
large-$T$ framework where $N/T\rightarrow c$ for some constant $c$ such that
$0<c<\infty$. As a natural way to model panel data with two-way effects, we
follow \cite{CHS_Restat} and assume that $y_{it}$ is generated as follows.

\begin{assumption}%
\begin{align*}
y_{it}=\theta+f\left(  {\alpha}_{i},{\gamma}_{t},{\varepsilon}_{it}\right)  ,
\end{align*}
where $\theta=\text{E}(y_{it})$ and $f$ is an unknown Borel-measurable
function, the sequences $\left\{  {\alpha}_{i}\right\}  $, $\left\{  {\gamma
}_{t}\right\}  $, and $\{{\varepsilon}_{it}\}$ are mutually independent,
${\alpha}_{i}$ is i.i.d across $i$, ${\varepsilon}_{it}$ is i.i.d across $i$
and $t$, and ${\gamma}_{t}$ is a strictly stationary serially correlated process.
\end{assumption}

The time component, ${\gamma}_{t}$, is allowed to have serial correlation
given that panel data typically has serial correlation beyond that induced by
individual effects. As pointed out by \cite{CHS_Restat} at the beginning of
their Section 3, the data-generating process in Assumption 1 is a strict
generalization of the representation developed by \cite{hoover1979relations},
\cite{aldous1981representations} and \cite{kallenberg1989representation} (the
so-called AHK representation) precisely because ${\gamma}_{t}$ is allowed to
have serial correlation. The AHK representation is not sufficient here because
it was developed for data drawn from an infinite array of jointly exchangeable
random variables in which case ${\gamma}_{t}$ would not have serial correlation.

Using the representation in Assumption 1, \cite{CHS_Restat} develop the
following decomposition of $y_{it}$. Denoting $a_{i}=\text{E}\left(
y_{it}-\theta|\alpha_{i}\right)  ,\ g_{t}=\text{E}\left(  y_{it}-\theta
|\gamma_{t}\right)  $, and $e_{it}=(y_{it}-\theta)-a_{i}-g_{t}$, one can
decompose $y_{it}-\theta$ as%
\[
y_{it}-\theta=a_{i}+g_{t}+e_{it}=: v_{it}.
\]
\cite{CHS_Restat} show that the components are mean zero and that $e_{it}$ is
also mean zero conditional on $a_{i}$ and conditional on $g_{t}$. The
individual component, $a_{i}$, is i.i.d.\ across $i$, and the time component,
$g_{t}$, is stationary. Conditional on $\gamma_{t}$, the $e_{it}$ component is
independent across $i$. Finally, the three components are uncorrelated with
each other with $a_{i}$ and $g_{t}$ being independent of each other. See
Section 3.1 of \cite{CHS_Restat} for details.

We can estimate $\theta$ using the pooled sample mean estimator given by 
$\widehat{\theta}={\left(  NT\right)  }^{-1}\sum_{i=1}^{N}{\sum_{t=1}%
^{T}{y_{it}}}$. Rewriting the sample mean using the component structure
representation for $y_{it}$ gives%
\begin{equation}
\widehat{\theta}-\theta=\frac{1}{N}\sum_{i=1}^{N}{a_{i}}+\frac{1}{T}\sum
_{t=1}^{T}{g_{t}}+\frac{1}{NT}\sum_{i=1}^{N}{\sum_{t=1}^{T}{e_{it}=:}%
}\text{ }\bar{a}+\bar{g}+\bar{e}. \label{thetahat}%
\end{equation}
The \cite{CHS_Restat} variance estimator of $\widehat{\theta}$ is given by (\ref{omega_chs}) with ${\hat{v}}%
_{it}=y_{it}-\widehat{\theta}$ used in (\ref{chs_a}) - (\ref{chs_nw}). To obtain
fixed-$b$ results for $\widehat{\Omega}_\text{CHS}$ we rewrite the formula for
$\widehat{\Omega}_\text{CHS}$ in terms of the following two partial sum processes
of ${\hat{v}}_{it}$:%
\begin{align}
{\widehat{S}}_{it}  &  =\sum_{j=1}^{t}{{\hat{v}}_{ij}}=t\left(  a_{i}%
-\bar{a}\right)  +\sum_{j=1}^{t}{\left(  g_{j}-\bar{g}\right)  }+\sum
_{j=1}^{t}{\left(  e_{ij}-\bar{e}\right)  ,}\label{Shati}\\
{\widehat{\bar{S}}}_{t}  &  =\sum_{i=1}^{N}{\widehat{S}}_{it}=\sum_{i=1}%
^{N}\sum_{j=1}^{t}{{\hat{v}}_{ij}}=N\sum_{j=1}^{t}{\left(  g_{j}-\bar
{g}\right)  }+\sum_{i=1}^{N}{\sum_{j=1}^{t}{\left(  e_{ij}-\bar{e}\right)  .}}
\label{Shatbar}%
\end{align}
Note that the $a_{i}$ component drops from (\ref{Shatbar}) because $\sum
_{i=1}^{N}{\left(  a_{i}-\bar{a}\right)  }=0$. The Arellano component
(\ref{chs_a}) of $\widehat{\Omega}_\text{CHS}$ is obviously a simple function of
(\ref{Shati}) with $t=T$. The HAC components (\ref{chs_dk}) and (\ref{chs_nw}) can
be written in terms of (\ref{Shatbar}) and (\ref{Shati}) using fixed-$b$
algebra (see \citealp{Vogelsang2012}). Therefore, the \cite{CHS_Restat} variance
estimator has the following equivalent formula:%
\begin{align}
\widehat{\Omega}_\text{CHS}=  &  \frac{1}{N^{2}T^{2}}\sum\limits_{i=1}^{N}%
\widehat{S}_{iT}\widehat{S}_{iT}^{\prime}\label{CHSps1}\\
&  +\frac{1}{N^{2}T^{2}}\left\{  \frac{2}{M}\sum\limits_{t=1}^{T-1}%
{{\widehat{\bar{S}}}}_{t}{\widehat{\bar{S}}}_{t}^{\prime}-\frac{1}{M}%
\sum\limits_{t=1}^{T-M-1}{\left(  {\widehat{\bar{S}}}_{t}{\widehat{\bar{S}}%
}_{t+M}^{\prime}+{\widehat{\bar{S}}}_{t+M}{\widehat{\bar{S}}}_{t}^{\prime
}\right)  }\right\} \label{CHSps2}\\
&  -\frac{1}{N^{2}T^{2}}\sum\limits_{i=1}^{N}\left\{  \frac{2}{M}%
\sum\limits_{t=1}^{T-1}{{\widehat{S}}_{it}{\widehat{S}}_{it}^{\prime}}%
-\frac{1}{M}\sum\limits_{t=1}^{T-M-1}{\left(  {\widehat{S}}_{it}{\widehat{S}%
}_{i,t+M}^{\prime}+{\widehat{S}}_{i,t+M}{\widehat{S}}_{i,t}^{\prime}\right)
}\right. \nonumber\\
&  \text{ \ \ \ \ \ \ \ \ \ \ \ \ \ \ \ \ \ \ \ \ }\left.  -\frac{1}{M}%
\sum\limits_{t=T-M}^{T-1}{\left(  {\widehat{S}}_{it}{\widehat{S}}_{iT}%
^{\prime}+{\widehat{S}}_{iT}{\widehat{S}}_{it}^{\prime}\right)  }+{\widehat
{S}}_{iT}{\widehat{S}}_{iT}^{\prime}\right\}  . \label{CHSps3}%
\end{align}
Define three $k\times k$ matrices $\Lambda_{a}$, $\Lambda_{g}$, and
$\Lambda_{e}$ such that:
\[
\Lambda_{a}\Lambda_{a}^{\prime}=\text{E}(a_{i}a_{i}^{\prime})\text{,
\ \ \ }\Lambda_{g}\Lambda_{g}^{\prime}=\sum_{\ell=-\infty}^{\infty}%
{\text{E}[g_{t}g_{t+\ell}^{\prime}]}\text{, \ \ \ }\Lambda_{e}\Lambda
_{e}^{\prime}=\sum_{\ell=-\infty}^{\infty}{\text{E}[e_{it}e_{i,t+\ell}%
^{\prime}].}%
\]
The following assumption is used to obtain an asymptotic result for
(\ref{thetahat}) and a fixed-$b$ asymptotic result for $\widehat{\Omega}%
_\text{CHS}$. Through out the paper, let $\Vert.\Vert$ denote the Euclidean norm
for matrices, and let $\lambda_{min}[.]$ denote the smallest eigenvalue of a
square matrix.

\begin{assumption}
For some $s>1$ and $\delta>0$, (i) $\emph{E}[y_{it}]=\theta$ and
$\emph{E}[\lVert y_{it}\rVert^{4(s+\delta)}]<\infty$. (ii) $\gamma_{t}$ is an
$\alpha$-mixing sequence with size $2s/(s-1)$, i.e., $\alpha_{\gamma}%
(\ell)=O(\ell^{-\lambda})$ for a $\lambda>2s/(s-1)$. (iii) $\lambda
_{min}[\Lambda_{a}\Lambda_{a}^{\prime}]>0$ and/or $\lambda_{min}[\Lambda
_{g}\Lambda_{g}^{\prime}]>0$, and $N/T\rightarrow c$ as $(N,T)\rightarrow
\infty$ for some constant $c$. (iv) $M=[bT]$ where $b\in(0,1]$.
\end{assumption}

Assumption 2(i) assumes the mean of $y_{it}$ exists and $y_{it}$ has finite
fourth moments. Assumption 2(ii) assumes weak dependence of $\gamma_{t}$ using
a mixing condition. Assumption 2 (i) - (ii) follow \cite{CHS_Restat}.
Assumption 2(iii) is a non-degeneracy restriction on the projected individual
and time components. Clearly, when data is i.i.d over both the cross-section
and time dimensions, this condition does not hold. Because the fixed-$b$
limits of $\widehat{\Omega}_\text{CHS}$ and its associated test statistics turn out
to be different in the i.i.d case, we discuss it separately in Section 4.
Assumption 2(iii) also rules out the pathological case described in Example
1.7 of \cite{Menzel2021}: when $y_{it}=\alpha_{i}\gamma_{t}+\varepsilon_{it}$
with $\text{E}(\alpha_{i})=\text{E}(\gamma_{t})=0$, one can easily verify that
$a_{i}=g_{t}=0$, in which case the limiting distribution of appropriately
scaled $\widehat{\theta}$ is non-Gaussian. Assumption 2(iv) uses the fixed-$b$
asymptotic nesting for the bandwidth. The following theorem gives an
asymptotic result for appropriately scaled $\widehat{\theta}$ and a fixed-$b$
asymptotic result for appropriately scaled $\widehat{\Omega}_\text{CHS}$.

\begin{theorem}
Let $z_{k}$ be a $k\times1$ vector of independent standard normal random
variables, and let $W_{k}(r)$, $r\in(0,1]$, be a $k\times1$ vector of
independent standard Wiener processes independent of $z_{k}$. Suppose
Assumptions 1 and 2 hold, then as $(N,T)\rightarrow\infty$,%
\begin{align}
    \sqrt{N}\left(  \widehat{\theta}-\theta\right) & \Rightarrow\Lambda_{a}%
z_{k}+\sqrt{c}\Lambda_{g}W_{k}(1), \nonumber \\
N\widehat{\Omega}_{\rm {CHS}} & \Rightarrow h(b)\Lambda_{a}\Lambda_{a}^{\prime
}+c\Lambda_{g}P\left(  b,{\widetilde{W}}_{k}\left(  r\right)  \right)
\Lambda_{g}^{\prime}, \label{omhatCHSlimit}%
\end{align}
where $h(b):=\left(  1-b+\frac{1}{3}b^{2}\right)$, $\widetilde{W}
_{k}(r):=W_{k}(r)-rW_{k}(1)$, and
\begin{align*}
P\left(  b,{\widetilde{W}}_{k}\left(  r\right)  \right) :=\frac{2}{b}\int
_{0}^{1}{{\widetilde{W}}}_{k}{\left(  r\right)  {\widetilde{W}}}_{k}{{\left(
r\right)  }^{\prime}dr}-\frac{1}{b}\int_{0}^{1-b}{\left[  {\widetilde{W}}%
_{k}\left(  r\right)  {\widetilde{W}}_{k}{\left(  r+b\right)  }^{\prime
}+{\widetilde{W}}_{k}\left(  r+b\right)  {\widetilde{W}}_{k}{\left(  r\right)
}^{\prime}\right]  }dr.
\end{align*}
\end{theorem}

The proof of Theorem 1 is given in the Appendix. The limit of $\sqrt{N}\left(
\widehat{\theta}-\theta\right)  $ was obtained by \cite{CHS_Restat}. Because
$z_{k}$ and $W_{k}(1)$ are vectors of independent standard normals that are
independent of each other, $\Lambda_{a}z_{k}+\sqrt{c}\Lambda_{g}W_{k}(1)$ is a
vector of normal random variables with variance-covariance matrix $\Lambda
_{a}\Lambda_{a}^{\prime}+c\Lambda_{g}\Lambda_{g}^{\prime}$. The $\Lambda
_{g}P\left(  b,{\widetilde{W}}\left(  r\right)  \right)  \Lambda_{g}^{\prime}$
component of (\ref{omhatCHSlimit}) is equivalent to the fixed-$b$ limit
obtained by \cite{Kiefer2005} in stationary time series settings. Obviously,
(\ref{omhatCHSlimit}) is different than the limit obtained by
\cite{Kiefer2005} because of the $h(b)\Lambda_{a}\Lambda_{a}^{\prime}$ term.
As the proof illustrates, this term is the limit of the \textquotedblleft
cluster by individuals\textquotedblright\ (\ref{chs_a}) and \textquotedblleft
average of HACs\textquotedblright\ (\ref{chs_nw}) components whereas the
$c\Lambda_{g}P\left(  b,{\widetilde{W}}_{k}\left(  r\right)  \right)
\Lambda_{g}^{\prime}$ term is the limit of the \textquotedblleft HAC of
averages\textquotedblright\ (\ref{chs_dk}). Interestingly, \cite{Kiefer2005}
showed that%
\begin{equation}
\text{E}\left(  P\left(  b,{\widetilde{W}}_{k}\left(  r\right)  \right)  \right)
=\left(  1-b+\frac{1}{3}b^{2}\right)  I_{k}=h(b)I_{k}, \label{meanPb}%
\end{equation}
where $I_{k}$ is a $k\times k$ identity matrix. The fact that both terms in
the limit of $N\widehat{\Omega}_\text{CHS}$ are proportional to $h(b)$ suggests a
simple bias correction that is discussed in Section 4.1. Because of the
component structure of (\ref{omhatCHSlimit}), the fixed-$b$ limits of $t$ and
Wald statistics based on $\widehat{\Omega}_\text{CHS}$ are not pivotal. We
provide details on test statistics after extending our results to the case of
a linear panel regression.

\subsection{The Linear Panel Regression Case}

\noindent It is straightforward to extend our results to the case of a linear
panel regression given by (\ref{reg}). The POLS estimator of $\beta$ is%
\begin{equation}
\widehat{\beta}=\beta+\widehat{Q}^{-1}\left(  \frac{1}{NT}\sum_{i=1}^{N}%
{\sum_{t=1}^{T}{x_{it}u_{it}}}\right),   \label{betahat}%
\end{equation}
where $\widehat{Q}:=\frac{1}{NT}\sum_{i=1}^{N}{\sum_{t=1}^{T}{x_{it}x_{it}^{\prime}}}$ as in Section 2. Following \cite{CHS_Restat}, we assume the components of the panel regression
are generated from the component structure:
\[
(y_{it},x_{it}^{\prime},u_{it})^{\prime}=f\left(  {\alpha}_{i},{\gamma}%
_{t},{\varepsilon}_{it}\right)
\]
where $f$ is an unknown Borel-measurable function, the sequences $\left\{
{\alpha}_{i}\right\}  $, $\left\{  {\gamma}_{t}\right\}  $, and
$\{{\varepsilon}_{it}\}$ are mutually independent, ${\alpha}_{i}$ is i.i.d
across $i$, ${\varepsilon}_{it}$ is i.i.d across $i$ and $t$, and ${\gamma
}_{t}$ is a strictly stationary serially correlated process. Define the vector
$v_{it}=x_{it}u_{it}$. Similar to the simple mean model we can write
$a_{i}=\text{E}\left(  v_{it}|a_{i}\right)  $, $g_{t}=\text{E}\left(
v_{it}|\gamma_{t}\right)  $, $e_{it}=v_{it}-a_{i}-g_{t}$, giving the
decomposition%
\[
v_{it}=a_{i}+g_{t}+e_{it}.
\]
The \cite{CHS_Restat} variance estimator of $\widehat{\beta}$ is given by (\ref{var_chs}) with ${\hat{v}}_{it}$ in (\ref{chs_a}) - (\ref{chs_nw})
now defined as ${\hat{v}}_{it}=x_{it}{\hat{u}}_{it}$ where
${\hat{u}}_{it}=y_{it}-x_{it}^{\prime}\widehat{\beta}$ are the POLS residuals.

The following assumption is used to obtain an asymptotic result for
(\ref{betahat}) and a fixed-$b$ asymptotic result for $\widehat{\Omega}_\text{CHS}$
in the linear panel case.

\begin{assumption}
For some $s>1$ and $\delta>0$, (i) $(y_{it},x_{it}^{\prime},u_{it})^{\prime
}=f(\alpha_{i},\gamma_{t},\varepsilon_{it})$ where $\{\alpha_{i}\}$,
$\{\gamma_{t}\}$, and $\{\varepsilon_{it}\}$ are mutually independent
sequences, $\alpha_{i}$ is i.i.d across $i$, $\varepsilon_{it}$ is i.i.d
across $i$ and $t$, and $\gamma_{t}$ is strictly stationary. (ii)
$\emph{E}[x_{it}u_{it}]=0$, $\lambda_{min}\left[  \emph{E}[x_{it}%
x_{it}^{\prime}]\right]  >0$, $\emph{E}[\lVert x_{it}\rVert^{8(s+\delta
)}]<\infty$, and $\emph{E}[\lVert u_{it}\rVert^{8(s+\delta)}]<\infty$. (iii)
$\gamma_{t}$ is an $\alpha$-mixing sequence with size $2s/(s-1)$, i.e.,
$\alpha_{\gamma}(\ell)=O(\ell^{-\lambda})$ for a $\lambda>2s/(s-1)$. (iv)
$\lambda_{min}[\Lambda_{a}\Lambda_{a}^{\prime}]>0$ and/or $\lambda
_{min}[\Lambda_{g}\Lambda_{g}^{\prime}]>0$, and $N/T\rightarrow c$ as
$(N,T)\rightarrow\infty$ for some constant $c$. (v) $M=[bT]$ where $b\in(0,1]$.
\end{assumption}

Assumption 3 can be regarded as a counterpart of Assumptions 1 and 2 with
Assumption 3(ii) being strengthened. It is very similar to its counterpart in
\cite{CHS_Restat} with a main difference the use of the fixed-$b$ asymptotic
nesting for the bandwidth, $M$. For the same reason mentioned in the previous
section, we discuss the case where $(x_{it},u_{it})$ are i.i.d separately in
Section 4.

The next theorem presents the joint limit of the POLS estimator and the
fixed-$b$ joint limit of CHS variance estimator.

\begin{theorem}
Let $z_{k}$, $W_{k}(r)$, ${\widetilde{W}}_{k}\left(  r\right)  $, $P\left(
b,{\widetilde{W}}_{k}\left(  r\right)  \right)  $ and $h(b)$ be defined as in
Theorem 1. Suppose Assumption 3 holds for model (\ref{reg}), then as
$(N,T)\rightarrow\infty$,
\[
\sqrt{N}\left(  \widehat{\beta}-\beta\right)  \Rightarrow Q^{-1}B_{k}(c),
\]
where $B_{k}(c):= \Lambda_{a}z_{k}+\sqrt{c}\Lambda_{g}W_{k}(1)$ and%
\begin{equation}
N\widehat{\emph{V}}_\emph{CHS}(\widehat{\beta})\Rightarrow Q^{-1}V_{k}%
(b,c)Q^{-1}, \label{NVhat}%
\end{equation}
where $V_{k}(b,c):= h(b)\Lambda_{a}\Lambda_{a}^{\prime}+c\Lambda
_{g}P\left(  b,{\widetilde{W}}_{k}(r)\right)  \Lambda_{g}^{\prime}.$
\end{theorem}

The proof of Theorem 2 is given in the Appendix. We can see that the limiting
random variable, $V_{k}(b,c)$, depends on the choice of truncation parameter,
$M$, through $b$. The use of the Bartlett kernel is reflected in the
functional form of $P\left(  b,{\widetilde{W}}_{k}(r)\right)  $ as well as the
scaling term $h(b)$ on $\Lambda_{a}\Lambda_{a}^{\prime}$. Use of a different
kernel would result in different functional forms for these limits. Because of
(\ref{meanPb}), it follows that%
\begin{align}
\text{E}\left(  V_{k}(b,c)\right)   &  =h(b)\Lambda_{a}\Lambda_{a}^{\prime
}+c\Lambda_{g}\text{E}\left[  P\left(  b,{\widetilde{W}}_{b}(r)\right)
\right]  \Lambda_{g}^{\prime}\nonumber\\
&  =h(b)\left(  \Lambda_{a}\Lambda_{a}^{\prime}+c\Lambda_{g}\Lambda
_{g}^{\prime}\right)  . \label{EVb}%
\end{align}
The scalar $h(b)$ can be viewed as a multiplicative bias term that depends on
the bandwidth sample size ratio, $b=M/T$. We leverage this fact to implement a
simple feasible bias correction for the CHS variance estimator that is
explored below.

Using the theoretical results developed in this section, we next examine the
properties of test statistics based on the POLS estimator and CHS variance
estimator. We also analyze tests based on two variants of the CHS variance
estimator. One is a bias-corrected estimator. The other is a variance
estimator guaranteed to be positive semi-definite that is also bias-corrected.

\section{Inference}

\noindent In regression model (\ref{reg}) we focus on tests of linear
hypothesis of the form:%
\[
\text{H}_{0}:R\beta=r,\text{ \ \ \ }\text{H}_{1}:R\beta\neq r,
\]
where $R$ is a $q\times k$ matrix ($q\leq k$) with full rank equal to $q$, and
$r$ is a $q\times1$ vector. Using $\widehat{\text{V}}_\text{CHS}(\widehat{\beta
})$ as given by (\ref{var_chs}), define a Wald statistic as%
\begin{align*}
W_\text{CHS}  &  ={\left(  R\widehat{\beta}-r\right)  }^{\prime}{\left(
R\widehat{\text{V}}_\text{CHS}(\widehat{\beta})R^{\prime}\right)  }^{-1}\left(
R\widehat{\beta}-r\right) .
\end{align*}
When $q=1$, we can define a $t$-statistic as%
\[
t_\text{CHS}=\frac{{R\widehat{\beta}-r}}{\sqrt{{R}\widehat{\text{V}}%
_\text{CHS}(\widehat{\beta})R^{\prime}}}.
\]
Appropriately scaling the numerators and denominators of the test statistics
and applying Theorem 2, we obtain under $\text{H}_{0}$:%
\begin{align}
W_\text{CHS}  &  =\sqrt{N}{\left(  R\widehat{\beta}-r\right)  }^{\prime}{\left(
RN\widehat{\text{V}}_\text{CHS}(\widehat{\beta})R^{\prime}\right)  }^{-1}\sqrt
{N}\left(  R\widehat{\beta}-r\right) \nonumber\\
&  \Rightarrow{\left(  RQ^{-1}B_{k}(c)\right)  }^{\prime}{\left(  RQ^{-1}%
V_{k}(b,c)Q^{-1}R^{\prime}\right)  }^{-1}\left(  RQ^{-1}B_{k}(c)\right) =: W_\text{CHS}^{\infty},\label{WaldCHSlimit}\\
t_\text{CHS}  &  =\frac{\sqrt{N}{\left(  R\widehat{\beta}-r\right)  }}{\sqrt
{{RN}\widehat{\text{V}}_\text{CHS}(\widehat{\beta})R^{\prime}}}\Rightarrow
\frac{RQ^{-1}B_{k}(c)}{\sqrt{{RQ^{-1}}V_{k}(b,c){Q^{-1}R^{\prime}}}}=:
t_\text{CHS}^{\infty}. \label{tCHSlimit}%
\end{align}
The limits of $W_\text{CHS}$ and $t_\text{CHS}$ are similar to the fixed-$b$ limits
obtained by \cite{Kiefer2005} but have distinct differences. First, the form
of $V_{k}(b,c)$ depends on two variance matrices rather than one. Second, the
variance matrices do not scale out of the statistics. Therefore, the fixed-$b$
limits given by (\ref{WaldCHSlimit}) and (\ref{tCHSlimit}) are not pivotal. We
propose a plug-in method for the simulation of critical values from these
asymptotic random variables.

For the case where $b$ is small, the fixed-$b$ critical values are close to
$\chi_{q}^{2}$ and $N(0,1)$ critical values respectively. This can be seen by
computing the probability limits of the asymptotic distributions as
$b\rightarrow0$. In particular, using the fact that $\text{plim}%
_{b\rightarrow0}P\left(  b,{\widetilde{W}}_{k}(r)\right)  =I_{k}$ (see
\citealp{Kiefer2005}), it follows that%
\begin{align*}
\text{plim}_{b\rightarrow0}V_{k}(b,c)  &  =\text{plim}_{b\rightarrow0}\left[
h(b)\Lambda_{a}\Lambda_{a}^{\prime}+c\Lambda_{g}P\left(  b,{\widetilde{W}}%
_{k}(r)\right)  \Lambda_{g}^{\prime}\right] \\
&  =\Lambda_{a}\Lambda_{a}^{\prime}+c\Lambda_{g}\Lambda_{g}^{\prime
}=\text{Var}(B_{k}(c)),
\end{align*}
where $h(\cdot)$ and $P(\cdot)$ are defined in Theorem 1. Therefore, it follows that%
\begin{align*}
&  \text{plim}_{b\rightarrow0}\left[  {\left(  RQ^{-1}B_{k}(c)\right)
}^{\prime}{\left(  RQ^{-1}V_{k}(b,c)Q^{-1}R^{\prime}\right)  }^{-1}\left(
RQ^{-1}B_{k}(c)\right)  \right] \\
&  ={\left(  RQ^{-1}B_{k}(c)\right)  }^{\prime}{\left(  RQ^{-1}\text{Var}%
(B_{k}(c))Q^{-1}R^{\prime}\right)  }^{-1}\left(  RQ^{-1}B_{k}(c)\right)
\sim\chi_{q}^{2},
\end{align*}
and
\[
p\lim_{b\rightarrow0}\left[  \frac{RQ^{-1}B_{k}(c)}{\sqrt{{RQ^{-1}}%
V_{k}(b,c){Q^{-1}R^{\prime}}}}\right]  =\frac{RQ^{-1}B_{k}(c)}{\sqrt
{{RQ^{-1}\text{Var}(}B_{k}(c){)Q^{-1}R^{\prime}}}}\sim N(0,1).
\]
In practice, there will not be a substantial difference between using
$\chi_{q}^{2}$ and $N(0,1)$ critical values and fixed-$b$ critical values for
small bandwidths. However, for larger bandwidths more reliable inference can
be obtained with fixed-$b$ critical values.

\subsection{Bias-Corrected CHS Variance Estimator}

We now leverage the form of the mean of the fixed-$b$ limit of the CHS
variance estimator as given by (\ref{EVb}) to propose a biased corrected
version of the CHS variance estimator. The idea is simple. We can scale out
the $h(b)$ multiplicative term evaluated at $b=M/T$ to make the CHS variance
estimator an asymptotically unbiased estimator of $\Lambda_{a}\Lambda
_{a}^{\prime}+c\Lambda_{g}\Lambda_{g}^{\prime}$, the variance of
$B_{k}(c)=\Lambda_{a}z_{k}+\sqrt{c}\Lambda_{g}W_{k}(1)$. Define the
bias-corrected CHS variance estimators as%
\begin{align*}
\widehat{\text{V}}_\text{BCCHS}\left(  \widehat{\beta}\right) ={\widehat{Q}%
}^{-1}{\widehat{\Omega}}_\text{BCCHS}{\widehat{Q}}^{-1}, \ \ {\widehat{\Omega}}_\text{BCCHS}  =\left(  h\left(  \frac{M}{T}\right)  \right)
^{-1}{\widehat{\Omega}}_\text{CHS},
\end{align*}
and the corresponding Wald and $t$-statistics under the null hypothesis
$R\beta=r$ are defined as
\begin{align*}
W_\text{BCCHS}  &  ={\left(  R\widehat{\beta}-r\right)  }^{\prime}{\left(
R\widehat{\text{V}}_\text{BCCHS}\left(  \widehat{\beta}\right)  R^{\prime
}\right)  }^{-1}\left(  R\widehat{\beta}-r\right) ,\\
t_\text{BCCHS}  &  =\frac{{R\widehat{\beta}-r}}{\sqrt{{R}\widehat{\text{V}%
}_\text{BCCHS}\left(  \widehat{\beta}\right)  R^{\prime}}}.
\end{align*}
Because ${\widehat{\Omega}}_\text{BCCHS}$ is a simple scalar multiple of
${\widehat{\Omega}}_\text{CHS}$, we easily obtain the fixed-$b$ limits
\begin{align}
W_\text{BCCHS}  &  \Rightarrow h(b)W_\text{CHS}^{\infty}=: W_\text{BCCHS}%
^{\infty},\label{WaldBCCHSlimit}\\
t_\text{BCCHS}  &  \Rightarrow h(b)^{1/2}t_\text{CHS}^{\infty}=: t_\text{BCCHS}^{\infty}.
\label{tBCCHSlimit}%
\end{align}

Notice that while the fixed-$b$ limits are different when using the
bias-corrected CHS variance estimator, they are scalar multiples of the
fixed-$b$ limits when using the original CHS variance estimator. Therefore,
the fixed-$b$ critical values of the $W_\text{BCCHS}$ and $t_\text{BCCHS}$ are
proportional to the fixed-$b$ critical values of $W_\text{CHS}$ and $t_\text{CHS}$.
As long as fixed-$b$ critical values are used, there is no practical effect on
inference from using the bias-corrected CHS variance estimator. Where the bias
correction matters is when $\chi_{q}^{2}$ and $N(0,1)$ critical values are
used. In this case, the bias-corrected CHS variance can provide more accurate
finite sample inference. This will be illustrated by our finite sample simulations.

\subsection{An Alternative Bias-Corrected Variance Estimator}

As noted by \cite{CHS_Restat}, the CHS variance estimator does not ensure
positive-definiteness, which is also the case for the clustered estimator
proposed by \cite{ColinCameron2011}. \cite{davezies2018asymptotic} and
\cite{mackinnon2021wild} point out that the double-counting adjustment term in
the estimator of \cite{ColinCameron2011} is of small order, and removing the
adjustment term has the computational advantage of guaranteeing positive
semi-definiteness. Analogously, we can think of ${\widehat{\Omega}}_\text{NW}$, as
given by (\ref{chs_nw}), as a double-counting adjustment term. If we exclude this
term, the variance estimator becomes the sum of two positive semi-definite
terms and is guaranteed to be positive definite. Another motivation for
dropping (\ref{chs_nw}) is that, under fixed-$b$ asymptotics, (\ref{chs_nw}) simply
contributes downward bias in the estimation of the $\Lambda_{a}\Lambda
_{a}^{\prime}$ term of $\text{Var}(B_{k}(c))$ through the $-b+\frac{1}{3}%
b^{2}$ part of $h(b)$ in the $h(b)\Lambda_{a}\Lambda_{a}^{\prime}$ portion of
$V_{k}(b,c)$. Intuitively, the Arellano cluster estimator takes care of the
serial correlation introduced by $a_{i}$, and the DK estimator takes care of
the cross-section and time dependence introduced by $g_{t}$. From this
perspective, ${\widehat{\Omega}}_\text{NW}$ is not needed.

Accordingly, we propose a variance estimator which is the sum of the Arellano
variance estimator and the bias-corrected DK variance estimator (labeled as
DKA hereafter) defined as%
\[
{\widehat{\Omega}}_\text{DKA}=:\widehat{\Omega}_\text{A}+h(b)^{-1}\widehat{\Omega}%
_\text{DK},
\]
where $\widehat{\Omega}_\text{A}$ and $\widehat{\Omega}_\text{DK}$ are defined in
(\ref{chs_a}) and (\ref{chs_dk}). Notice that we bias correct the DK component so
that the resulting variance estimator is asymptotically unbiased under
fixed-$b$ asymptotics. This can improve inference should $\chi_{q}^{2}$ or
$N(0,1)$ critical values be used in practice. The following theorem gives the
fixed-$b$ limit of the scaled DKA variance estimator.

\begin{theorem}
Suppose Assumption 3 holds for model (\ref{reg}), then as $(N,T)\rightarrow
\infty$,
\begin{equation}
N{\widehat{\Omega}}_\emph{DKA}\Rightarrow\Lambda_{a}\Lambda_{a}^{\prime}%
+c\Lambda_{g}h(b)^{-1}P\left(  b,{\widetilde{W}}_{b}\left(  r\right)  \right)
\Lambda_{g}^{\prime}=h(b)^{-1}V_{k}(b,c). \label{OmhatDKAlimit}%
\end{equation}

\end{theorem}

\noindent The proof of Theorem 3 can be found in the Appendix. Define the
statistics $W_\text{DKA}$ and $t_\text{DKA}$ analogous to $W_\text{BCCHS}$ and
$t_\text{BCCHS}$ using the variance estimator for $\widehat{\beta}$ given by $\widehat{\text{V}}_\text{DKA}\left(  \widehat{\beta}\right)  ={\widehat{Q}}%
^{-1}{\widehat{\Omega}}_\text{DKA}{\widehat{Q}}^{-1}$. Applying Theorems 2 and 3, we obtain the fixed-$b$ limits of Wald/t test
statistics associated with DKA variance estimator under the null:
\[
W_\text{DKA}\Rightarrow W_\text{BCCHS}^{\infty},\text{ \ \ }t_\text{DKA}\Rightarrow
t_\text{BCCHS}^{\infty},
\]
which are the same as the limits of $W_\text{BCCHS}$ and $t_\text{BCCHS}$ given by
(\ref{WaldBCCHSlimit}) and (\ref{tBCCHSlimit}).

\subsection{Results for i.i.d.\ Data}

While the DKA variance estimator is guaranteed to be positive semi-definite,
this useful property comes with a potential cost. As is shown in Theorem 2 of
\cite{mackinnon2021wild}, if the score $x_{it}u_{it}$ is i.i.d. over $i$ and
$t$, or if clusters are formed at the intersection between individuals and
time\footnote{In our setting where the clustering only happens at individual
and time levels, clustering at the intersection is the same as the
independence across individuals and times.}, the probability limit of two-way cluster-robust variance
estimators that drop the double-counting adjustment term, referred to as a
two-term variance estimator, is twice the size of the true variance. In other
words, if the researcher believes there is clustering when there is none, the
use of a two-term estimator would overestimate the asymptotic variance. The
associated Wald and $t$-statistics will be scaled down causing over-coverage
(under-rejection) problems under the null hypothesis. The following assumption
and theorem give fixed-$b$ results for the CHS, BCCHS and DKA statistics for
the case of i.i.d. data.

\begin{assumption}
For some $s>1$ and $\delta>0$, (i) $(x_{it},u_{it})$ are independent and
identically distributed over i and t. (ii) $\text{E}[x_{it}u_{it}]=0$,
$\lambda_{min}\left[  \text{E}[x_{it}x_{it}^{\prime}]\right]  >0$,
$\text{E}[\lVert x_{it} \rVert^{8(s+\delta)}]<\infty$, and $\text{E}[\lVert
u_{it} \rVert^{8(s+\delta)}]<\infty$. (iii) $N/T \to c$ as $(N,T)\to\infty$
for some constant $c$. (iv) $M=[bT]$ for $b\in(0,1]$.
\end{assumption}



\begin{theorem}
Suppose Assumption 4 holds for model (\ref{reg}), then as $(N,T)\rightarrow
\infty,$%
\begin{align*}
W_{\emph{CHS}}  &  \Rightarrow W_{q}(1)^{\prime}\left\{  P\left(  b,\widetilde
{W}_{q}(r)\right)  \right\}  ^{-1}W_{q}(1) =: W_\emph{CHS}^{\infty,iid}, \ \  W_{\emph{BCCHS}}   \Rightarrow h(b)W_\emph{CHS}^{\infty,iid} =: W_{\emph{BCCHS}}%
^{\infty,iid},\\
W_\emph{DKA}  &  \Rightarrow W_{q}(1)^{\prime}\left\{  I_{q}+h(b)^{-1}P\left(
b,\widetilde{W}_{q}(r)\right)  \right\}  ^{-1}W_{q}(1)=: W_\emph{DKA}%
^{\infty,iid}, \\
t_\emph{CHS}  &  \Rightarrow\frac{W_{1}(1)}{\sqrt{P\left(  b,\widetilde{W}%
_{1}(r)\right)  }}=: t_\emph{CHS}^{\infty,iid}, \ \ t_{\emph{BCCHS}}\Rightarrow
h(b)^{1/2}t_\emph{CHS}^{\infty,iid}=: t_{\emph{BCCHS}}^{\infty,iid},\\
t_\emph{DKA}  &  \Rightarrow\frac{W_{1}(1)}{\sqrt{1+h(b)^{-1}P\left(
b,\widetilde{W}_{1}(r)\right)  }}=: t_\emph{DKA}^{\infty,iid}.
\end{align*}

\end{theorem}

\noindent Theorem 4 shows that the fixed-$b$ limits in the i.i.d. case are
different for all three test statistics than the limits given by
(\ref{WaldCHSlimit}), (\ref{tCHSlimit}) for CHS and (\ref{WaldBCCHSlimit}),
(\ref{tBCCHSlimit}) for BCCHS and DKA.

Suppose tests are carried out using $\chi_{q}^{2}$ \ and $N(0,1)$ critical
values. The limits in Theorem 4 can be used to compute asymptotic null
rejection probabilities, or equivalently, asymptotic coverage probabilities
for the case of i.i.d. data. For a two-tailed 5\% $t$-test, the coverage
probabilities are given by%
\[
P\left(  \left\vert t_\text{CHS}^{\infty,iid}\right\vert \leq1.96\right)  ,\text{
\ }P\left(  \left\vert t_\text{BCCHS}^{\infty,iid}\right\vert \leq1.96\right)
,\text{ \ }P\left(  \left\vert t_\text{DKA}^{\infty,iid}\right\vert \leq
1.96\right)  .
\]
For small bandwidths, we can analytically compute these asymptotic coverage
probabilities. As $b\rightarrow0$, the limits of $t_\text{CHS}^{\infty,iid}$ and
$t_\text{BCCHS}^{\infty,iid}$ converge to $N(0,1)$ random variables giving
asymptotic coverage of 95\%. In contrast as $b\rightarrow0$, the limit of
$t_\text{DKA}^{\infty,iid}$ is a $N(0,\frac{1}{2})$ random variable and the
asymptotic coverage is 99.4\%, and DKA over-covers and is conservative. This
result for DKA tests is similar to Corollary 1 of \cite{mackinnon2021wild}.
For non-small bandwidths the limiting random variables are non-standard. We
used simulation methods to compute these probabilities. We approximated the
Wiener processes using scaled partial sums of 1,000 i.i.d. $N(0,1)$ random
increments and used 50,000 replications to simulate the percentiles.

\begin{table}[th]
\label{tabblechsa} \centering{\small \begin{threeparttable}
\begin{tabular}
[c]{c|cccc|ccccc}%
\multicolumn{10}{c}{Table 4.1: Asymptotic Critical Values and Coverage
Probabilities (\%)}\\\hline\hline
&  & \multicolumn{1}{c}{} & \multicolumn{1}{c}{} & \multicolumn{1}{c|}{} &
\multicolumn{5}{c}{Coverage, $N(0,1)$ \& $\hat{t}_\text{BCCHS}^{\infty,iid}$ Critical Values}\\
& \multicolumn{4}{|c|}{97.5\% Asymptotic Critical Values} & CHS &
\multicolumn{2}{c}{BCCHS} & \multicolumn{2}{c}{DKA}\\
$b$ & $t_\text{CHS}^{\infty,iid}$ & $t_\text{BCCHS}^{\infty,iid}$ & $t_\text{DKA}%
^{\infty,iid}$ & $\hat{t}_\text{BCCHS}^{\infty,iid}$ & $N(0,1)$ & $N(0,1)$ & $\hat{t}_\text{BCCHS}^{\infty,iid}$ &
$N(0,1)$ & $\hat{t}_\text{BCCHS}^{\infty,iid}$\\\hline
0 & 1.960 & 1.960 & 1.386 & 1.960 & 95.0 & 95.0 & 95.0 & 99.4 & 99.4\\
0.08 & 2.191 & 2.104 & 1.411 & 1.972 & 92.5 & 93.5 & 93.7 & 99.4 & 99.4\\
0.12 & 2.298 & 2.162 & 1.416 & 1.991 & 91.2 & 92.8 & 93.2 & 99.3 & 99.4\\
0.16 & 2.421 & 2.230 & 1.425 & 2.006 & 89.8 & 92.2 & 92.7 & 99.2 & 99.3\\
0.20 & 2.546 & 2.296 & 1.438 & 2.019 & 88.6 & 91.5 & 92.3 & 99.1 & 99.3\\
0.40 & 3.181 & 2.571 & 1.470 & 2.070 & 82.2 & 89.0 & 90.5 & 98.9 & 99.3\\
0.80 & 4.300 & 2.764 & 1.497 & 2.100 & 71.2 & 87.3 & 89.2 & 98.8 & 99.2\\
1.00 & 4.791 & 2.766 & 1.497 & 2.099 & 66.7 & 87.2 & 89.2 & 98.8 &
99.2\\\hline
\end{tabular}
\begin{tablenotes}
\small
\item Note: Asymptotic critical values and coverage probabilities based on 50,000 replications
using 1,000 increments for the Wiener process. The random variables $\hat{t}_\text{BCCHS}^{\infty,iid}$ and $\hat{t}_\text{DKA}^{\infty,iid}$ are the same. The nominal coverage probability is 95\%.
\end{tablenotes}
\end{threeparttable}}\end{table}

Table 4.1 reports 97.5\% critical values for $t_\text{CHS}^{\infty,iid}$,
$t_\text{BCCHS}^{\infty,iid}$, and $t_\text{DKA}^{\infty,iid}$ for a range of values of
$b$ that will be used in our finite sample simulations. The critical values of
$t_\text{CHS}^{\infty,iid}$ and $t_\text{BCCHS}^{\infty,iid}$ equal 1.96 when $b=0$ and
increase as $b$ increase. This suggests that CHS and BCCHS tests will
under-cover when the data is i.i.d. and bandwidths are not small. In contrast,
the critical values of $t_\text{DKA}^{\infty,iid}$ are always smaller than 1.96 and
remain smaller as $b$ increases. Thus, DKA tests over-cover regardless of the bandwidth.

Table 4.1\ also reports asymptotic coverage probabilities using the $N(0,1)$
critical value. We see that as $b$ goes from $0$ to $1.0$, coverage decreases
from 95\% to 66.7\% for CHS, 95\% to 87.2\% for BCCHS, and is always close to
99\% for DKA. These asymptotic calculations predict that CHS and BCCHS will
over-reject (be liberal) when data is i.i.d. and non-small bandwidths are
used. DKA is predicted to be conservative regardless of bandwidth. The table
also reports some results for a random variable, $\hat{t}_\text{BCCHS}%
^{\infty,iid}$, that is discussed in the next section.

\subsection{Simulated fixed-$b$ Critical Values}

\noindent As we have noted, the fixed-$b$ limits of the test statistics given
by (\ref{WaldCHSlimit}), (\ref{tCHSlimit}) and (\ref{WaldBCCHSlimit}),
(\ref{tBCCHSlimit}) are not pivotal due to the nuisance parameters
$\Lambda_{a}$ and $\Lambda_{g}$. A feasible method for obtaining asymptotic
critical values is to use simulation methods with unknown nuisance parameters
replaced with estimators, i.e. use a plug-in simulation method.

To estimate $\Lambda_{a}$ and $\Lambda_{g}$ we use the estimators:
\begin{flalign*}
\widehat{{\Lambda}_a{\Lambda}'_a} & := \frac{1}{NT^2}\sum^N_{i=1}{\left(\sum^T_{t=1}{{\hat{v}}^{\ }_{it}}\right)\left(\sum^T_{s=1}{{\hat{v}}'_{is}}\right)},  \\
\widehat{{\Lambda}_g{\Lambda}'_g} & := {\left(1-b_\text{dk}+\frac{1}{3}b^2_\text{dk}\right)}^{-1}\frac{1}{N^2T}\sum^T_{t=1}{\sum^T_{s=1}{k\left(\frac{\left|t-s\right|}{M_\text{dk}}\right)\left(\sum^N_{i=1}{{\hat{v}}_{it}}\right)\left(\sum^N_{j=1}{{\hat{v}}'_{js}}\right)}}.
\end{flalign*}\noindent where $b_\text{dk}=\frac{M_\text{dk}}{T}$ and $M_\text{dk}$ is the
truncation parameter for the Driscoll-Kraay variance estimator.\footnote{Note
that, in principle, $b_\text{dk}$ can be different from the $b$ used for CHS
variance estimator. For simulating asymptotic critical values we used the data
dependent rule of Andrews (1991) to obtain $b_\text{dk}$.} The consistency of
$\widehat{\Lambda_{a}\Lambda_{a}^{\prime}}$ is given by (\ref{lambdaA}) in the
proof of Theorem 3:
\begin{equation}
\widehat{\Lambda_{a}\Lambda_{a}^{\prime}}=\Lambda_{a}\Lambda_{a}^{\prime
}+o_{p}(1), \label{lambda_ah}%
\end{equation}
And by (\ref{A.10}) in the proof of Theorem 2, we have,
\begin{equation}
\widehat{{\Lambda}_{g}{\Lambda}_{g}^{\prime}}\Rightarrow
{\Lambda}_{g}\frac{P\left(  b_\text{dk},{\widetilde{W}}\left(  r\right)
\right)  }{1-b_\text{dk}+\frac{1}{3}b_\text{dk}^{2}}{\Lambda}_{g}^{\prime}.
\label{lambda_gh}%
\end{equation}

Therefore, $\widehat{{\Lambda}_{a}{\Lambda}_{a}^{\prime}}$
is a consistent estimator for ${\Lambda}_{a}{\Lambda}%
_{a}^{\prime}$ and $\widehat{{\Lambda}_{g}{\Lambda}%
_{g}^{\prime}}$ is a bias-corrected estimator of ${\Lambda}%
_{g}{\Lambda}_{g}^{\prime}$ with the mean of the limit equal to
${\Lambda}_{g}{\Lambda}_{g}^{\prime}$ and the limit
converges to ${\Lambda}_{g}{\Lambda}_{g}^{\prime}$ as
$b_\text{dk}\rightarrow0$. The matrices ${\widehat{\Lambda}}_{a}$ and
${\widehat{\Lambda}}_{g}$ are matrix square roots of $\widehat
{{\Lambda}_{a}{\Lambda}_{a}^{\prime}}$ and $\widehat
{{\Lambda}_{g}{\Lambda}_{g}^{\prime}}$ respectively such
${\widehat{\Lambda}}_{a}{\widehat{\Lambda}}_{a}^{\prime
}=\widehat{{\Lambda}_{a}{\Lambda}_{a}^{\prime}}$ and
${\widehat{\Lambda}}_{g}{\widehat{\Lambda}}_{g}^{\prime
}=\widehat{{\Lambda}_{g}{\Lambda}_{g}^{\prime}}$

We propose the following plug-in method for simulating the asymptotic critical
values of the fixed-$b$ limits. Details are given for a $t$-test with the
modifications needed for a Wald test being obvious.

\begin{enumerate}
\item For a given data set with sample sizes $N$ and $T$, calculate
$\widehat{Q}$, ${\widehat{\Lambda}}_{a}$ and ${\widehat{\Lambda}}_{g}$. Let
$b=M/T$ where $M$ is the bandwidth used for $\widehat{\Omega}_\text{CHS}$. Let
$c=N/T.$

\item Taking $\widehat{Q}$, ${\widehat{\Lambda}}_{a}$, ${\widehat{\Lambda}%
}_{g}$, $b$, $c$, and $R$ as given, use Monte Carlo methods to simulate
critical values for the distributions%
\begin{align}
\hat{t}_\text{CHS}  &  =\frac{RQ^{-1}\hat{b}_{k}\left(  c\right)  }%
{\sqrt{{RQ^{-1}\hat{v}_{k}(b,c)Q^{-1}R^{\prime}}}}\label{tchshat}\\
\hat{t}_\text{BCCHS}  &  =\hat{t}_\text{DKA}=h(b)^{1/2}\hat{t}_\text{CHS}
\label{tdkahat}%
\end{align}
where
\begin{align*}
\hat{b}_{k}\left(  c\right)   &  =R\widehat{Q}^{-1}\left(  {\widehat
{\Lambda}}_{a}z_{k}+\sqrt{c}{\widehat{\Lambda}}_{g}W_{k}(1)\right)  ,\\
\hat{v}_{k}(b,c)  &  =h(b){\widehat{\Lambda}}_{a}{\widehat{\Lambda}}%
_{a}^{\prime}+c\widehat{\Lambda}_{g}P\left(  b,{\widetilde{W}}_{k}(r)\right)
\widehat{\Lambda}_{g}^{\prime}.
\end{align*}

\item Typically the process ${W}_{k}(r)$ is approximated using scaled partial
sums of a large number of i.i.d.\ $N(0,I_{k})$ realizations\ (increments) for
each replication of the Monte Carlo simulation.
\end{enumerate}

It is clear that under Assumption 3, as $N,T\rightarrow\infty$ and
then as $b_\text{dk}\rightarrow0$, $\hat{t}_\text{CHS}$ converges weakly to the
fixed-$b$ limit of $t_\text{CHS}$ in (\ref{WaldCHSlimit}) using results in
(\ref{lambda_ah}) and (\ref{lambda_gh}); and so does $\hat{t}_\text{BCCHS}$
($\hat{t}_\text{DKA}$). However, it is less clear what $\hat{t}_\text{CHS}$ and $\hat{t}_\text{BCCHS}$
($\hat{t}_\text{DKA}$) are estimating when data is i.i.d.\ across both
individual and time dimensions. In the i.i.d.\ case $\frac{1}{N}%
\widehat{\Lambda_{a}\Lambda_{a}^{\prime}}$ and $\frac{1}{T}\widehat
{{\Lambda}_{g}{\Lambda}_{g}^{\prime}}$ each estimate
$\Lambda_{xu}\Lambda_{xu}^{\prime}$ (the variance of $x_{it}u_{ut}$). Treating
$\frac{1}{N}\widehat{\Lambda_{a}\Lambda_{a}^{\prime}}$ , $\frac{1}{T}%
\widehat{{\Lambda}_{g}{\Lambda}_{g}^{\prime}}$, and
$\widehat{Q}$ as consistent plug-in estimators, it can be shown using
arguments similar to the proof of Theorem 4 that $\hat{t}_\text{CHS}$ and
$\hat{t}_\text{BCCHS}$ ($\hat{t}_\text{DKA}$) are simulating from the random
variables%
\[
\hat{t}_\text{CHS}^{\infty,iid}=\frac{z_{1}+W_{1}(1)}{\sqrt{h(b)+P\left(
b,\widetilde{W}_{1}(r)\right)  }},\text{ \ \ }\hat{t}_\text{BCCHS}^{\infty
,iid}=\hat{t}_\text{DKA}^{\infty,iid}=h(b)^{1/2}\hat{t}_\text{CHS}^{\infty
,iid}.
\]
While these random variables do not depend on $c$ or nuisance parameters, they
are clearly different than the limits given by Theorem 4. If we take the
probability limit of these random variables as $b\rightarrow0$, it is easy to
see\footnote{Obviously $\left(  1-b+\frac{1}{3}b^{2}\right)  \rightarrow1$ as
$b\rightarrow0$, and recall that $p\lim_{b\rightarrow0}P\left(  b,\widetilde
{W}_{q}(r)\right)  =I_{q}$.} that both random variables converge to%
\[
\frac{z_{1}+W_{1}(1)}{\sqrt{2}}\sim N(0,1),
\]
because $\left(  z_{1}+W_{1}(1)\right)  \sim N(0,2)$. Recall from Theorem 4
that the fixed-$b$ limits of $t_\text{CHS}$ and $t_\text{BCCHS}$ are also approximately
$N(0,1)$ when $b$ is small. Thus, the simulated critical values for $t_\text{CHS}$
and $t_\text{BCCHS}$ adapt to the i.i.d.\ case at least for small bandwidths. In
contrast, the limit of $t_\text{DKA}$ in Theorem 4 is approximately $N(0,\frac
{1}{2})$ when $b$ is small whereas $\hat{t}_\text{DKA}$ is simulating from a
$N(0,1)$ random variable. Therefore, simulated critical values for $t_\text{DKA}$
do not adapt to i.i.d.\ data when $b$ is small and $t_\text{DKA}$ over-covers and
is conservative.

When the plug-in critical values are used, we can make theoretical predictions
for coverage probabilities in the i.i.d. case for bandwidths that are not
small ($b>0$) by computing coverage probabilities of the limiting random
variables $t_\text{BCCHS}^{\infty,iid}$ and $t_\text{DKA}^{\infty,iid}$ using critical
values from the asymptotic random variable $\hat{t}_\text{BCCHS}^{\infty,iid}$
(same as $\hat{t}_\text{DKA}^{\infty,iid}$)\footnote{Coverage probabilities are
the same for $t_\text{CHS}^{\infty,iid}$ and $t_\text{BCCHS}^{\infty,iid}$ using
critical values from $\hat{t}_\text{CHS}^{\infty,iid}$ and $\hat{t}%
_\text{BCCHS}^{\infty,iid}$ given the common scaling factor $h(b)^{1/2}$.}. Results
are given in Table 4.1 in the $\hat{t}_\text{BCCHS}^{\infty,iid}$ columns. As
shown in the table, the critical values of $\hat{t}_\text{BCCHS}^{\infty,iid}$
increase with $b$ but slowly. This helps reduce the under-rejection problems
of BCCHS but does not remove them as we see in the coverage probability column
for BCCHS that uses critical values from $\hat{t}_\text{BCCHS}^{\infty,iid}$.
When DKA uses critical values from $\hat{t}_\text{BCCHS}^{\infty,iid}$,
coverage probabilities are similar to the $N(0,1)$, and the coverages do not
vary much across $b$ because the critical values of $t_\text{DKA}^{\infty,iid}$ and
$\hat{t}_\text{BCCHS}^{\infty,iid}$ roughly move together as $b$ increases.

The asymptotic calculations in Table 4.1 predict that CHS and BCCHS will tend
to under-cover (liberal) when the data is i.i.d. with the coverage approaching
the nominal level for small bandwidths. DKA is predicted to have over-coverage
(conservative) when the data is i.i.d. regardless of the bandwidth.

\section{Monte Carlo Simulations}

To illustrate the finite sample performance of the various variance estimators
and corresponding test statistics, we present a Monte Carlo simulation study
with 10,000 replications in all cases. We focus on a simple linear panel
model:
\begin{equation}
y_{it}={\beta}_{0}+{\beta}_{1}x_{it}+u_{it},\label{simple}%
\end{equation}
where  the true parameters are $(\beta_{0},\beta_{1})=(1,1)$. To allow direct
comparisons with Table 1 of \cite{CHS_working}, we consider a data generating
process (DGP) that is linear in the components:
\begin{align*}
\text{DGP}(1):\ ~x_{it} &  ={\omega}_{\alpha}{\alpha}_{i}^{x}+{\omega}_{\gamma
}{\gamma}_{t}^{x}+{\omega}_{\varepsilon}{\varepsilon}_{it}^{x},\\
u_{it} &  ={\omega}_{\alpha}{\alpha}_{i}^{u}+{\omega}_{\gamma}{\gamma}_{t}%
^{u}+{\omega}_{\varepsilon}{\varepsilon}_{it}^{u},\\
\gamma_{t}^{(j)} &  =\rho_{\gamma}{\gamma}_{t-1}^{(j)}+{\widetilde{\gamma}%
}_{t}^{(j)}\text{ for }j=x,u\text{,}%
\end{align*}
where the latent components $\{{\alpha}_{i}^{x},{\alpha}_{i}^{u},{\varepsilon
}_{it}^{x},{\varepsilon}_{it}^{u}\}$ are each i.i.d $N(0,1)$, and the error
terms $\widetilde{\gamma}_{t}^{(j)}$ for the AR(1) processes are i.i.d
$N(0,1-\rho_{\gamma}^{2})$ for $j=x,u$. The component weights $(\omega_\alpha,\omega_\gamma,\omega_\varepsilon)$ are used to adjust the relative importance of those components. 

To further explore the role played by the component structure representation,
we consider a second DGP where the latent components enter $x_{it}$ and
$u_{it}$ in a non-linear way:
\begin{align*}
\text{DGP}(2):\ ~x_{it}  &  =log(p_{it}^{(x)}/(1-p_{it}^{(x)})),\\
u_{it}  &  =log(p_{it}^{(u)}/(1-p_{it}^{(u)})),\\
p_{it}^{(j)}  &  =\Phi({\omega}_{\alpha}{\alpha}_{i}^{(j)}+{\omega}_{\gamma
}{\gamma}_{t}^{(j)}+{\omega}_{\varepsilon}{\varepsilon}_{it}^{(j)})\text{ for
}j=x,u\text{,}%
\end{align*}
where $\Phi(\cdot)$ is the cumulative distribution function of a standard
normal distribution and the latent components are generated in the same way as DGP(1).

Sample coverage probabilities of 95\% confidence intervals for $\widehat
{{\beta}}_{1}$, the OLS estimator of the slope parameter from (\ref{simple}),
are provided for the following variance estimators: Eicker-Huber-White (EHW),
cluster-by-$i$ (Ci), cluster-by-$t$ (Ct), DK, CHS, BCCHS, and DKA. For the
variance estimators that require a bandwidth choice (DK, CHS, BCCHS, and DKA)
we report results using the \cite{Andrews1991} AR(1) plug-in data-dependent
bandwidth, labeled as $\hat{M}$, designed to minimize the approximate mean
square error of a variance estimator (same formula for all four variance
estimators). In the case of a scalar $x_{it}$, the formula is given
by\footnote{Using equation (6.4) from \cite{Andrews1991}, we use 0 weight for
constant regressor and the weights equal to the inverse of the squared
innovation variances for other regressors. Because \cite{CHS_Restat} parameterize the Bartlett kernel as $1-\frac{m}{M+1}$ whereas we use $1-\frac{m}{M}$, we add 1 to the data-dependent formula so that our Bartlett weights match those used by \cite{CHS_Restat}.}
\[
\hat{M}=1.8171\left(  \frac{\widehat{\rho}^{2}}{\left(  1-\widehat{\rho
}^{2}\right)  ^{2}}\right)  ^{1/3}T^{1/3} + 1,
\]
where $\widehat{\rho}$ is the OLS estimator from the regression $\bar
{\hat{v}}_{t}=\rho\bar{\hat{v}}_{t-1}+\eta_{t}$ where
$\bar{\hat{v}}_{t}=\frac{1}{N}\sum_{i=1}^{N}\hat{v}_{it}$,
$\hat{v}_{it}=x_{it}\hat{u}_{it}$, and $\hat{u}_{it}$ are the OLS
residuals from (\ref{simple}). We label the ratio of $\hat{M}$ relative to
the time sample size as $\hat{b}=\hat{M}/T$. In some cases
$\hat{M}$ can exceed $T$ especially when the time dependence is strong
relative to $T$. Therefore, we truncate $\hat{M}$ at $T$ whenever
$\hat{M}>T$. We also report results for a grid of bandwidth choices. For
tests based on CHS and DKA, we use both the standard normal critical values
and the plug-in fixed-$b$ critical values. The simulated critical values use
1000 replications with the Wiener process approximated by scaled partial sums
of 500 independent increments drawn from a standard normal distribution. While
these are relatively small numbers of replications and increments for an
asymptotic critical value simulation, it was necessitated by computational
considerations given the need to run an asymptotic critical value simulation
for \textit{each} replication of the finite sample simulation.

\subsection{Main Simulation Results}

We first focus on DGP(1) to make direct comparisons to the simulation results
of Table 1 of \cite{CHS_working}, a working paper version of \cite{CHS_Restat}\footnote{The reason we refer to the 2022 working paper version of \cite{CHS_Restat} is because their results for small sample sizes $(N,T) = (25,25)$ are not included in the published paper, \cite{CHS_Restat}.}. Empirical null coverage probabilities of the
confidence intervals for ${\widehat{\beta}}_{1}$ are presented in Table
\ref{tablechs}. We start with both\ the cross-section and time sample sizes
equal to 25. The weights on the latent components are ${\omega}_{\alpha}%
=0.25$, ${\omega}_{\gamma}=0.5$, ${\omega}_{\varepsilon}=0.25$. Because of the
relatively large weight on the common time effect, $\gamma_{t}$, the
cross-section dependence dominates the time dependence. We can see that the
confidence intervals using EHW, Ci, and Ct \footnote{Finite sample adjustments
are applied to these three variance estimators. $\text{HC}_{1}$ is used for
EHW estimator. The \textquotedblleft cluster-by-$i$\textquotedblright, and
\textquotedblleft cluster-by-$t$\textquotedblright\ are also adjusted by the
usual degrees-of-freedom factor.} suffer from a severe under-coverage problems
as they fail to capture both cross-section and time dependence.

\begin{table}[th]
\label{tablechs} \centering{\begin{threeparttable}
\begin{tabular}
[c]{cc|cc|cccccc|c}%
\multicolumn{11}{c}{Table 5.1: Sample Coverage Probabilities (\%), Nominal Coverage
95\%}\\
\multicolumn{10}{c}{$N=T=25$; DGP(1): $\omega_{\alpha}=\omega
_{\varepsilon}=0.25$, $\omega_{\gamma}=0.5$, $\rho_{\gamma}= 0.425$; POLS.}\\\hline\hline
&  &  &  &  &  & BC- &  & \multicolumn{2}{c|}{fixed-b c.v.} & Count \\\cline{9-10}
\multicolumn{2}{c|}{No Truncation}&$M$ & $b$ & DK & CHS & CHS & DKA & CHS & DKA & CHS$<0$ \\\hline
EHW& 37.4&$\hat{M}$&$\hat{b}$&83.6&84.1& 86.2&88.1& 88.3&90.3& 0\\
Ci & 38.7&2  & 0.08                  &84.0&84.4& 86.0&87.9& 88.1&89.7& 0\\
Ct & 83.6&3  & 0.12                  &83.3&83.7& 85.8&87.9& 87.8&90.1& 0\\
&     &4  & 0.16                  &82.0&82.3& 85.3&87.5& 88.2&90.5& 0\\
&     &5  & 0.20                  &80.6&80.8& 84.8&87.3& 88.1&90.3& 0\\
&     &10 & 0.40                  &73.9&74.4& 82.2&85.4& 88.5&91.2& 0\\
&     &20 & 0.80                  &62.6&63.0& 80.9&84.0& 87.9&91.1& 0\\
&     &25 & 1.00                  &57.9&58.4& 80.8&84.0& 88.2&90.9& 0\\ \hline
\end{tabular}
\begin{tablenotes}
\small
\item Note: $\hat{M}$ ranged from 1 to 21, with an average of 2.6 and median of 2.
\end{tablenotes}
\end{threeparttable}}\end{table}

With the time effect, $\gamma_{t}$, being mildly persistent ($\rho_{\gamma
}=0.425$), the DK and CHS confidence intervals using the normal approximation
undercover with small bandwidths with empirical rejection rates mostly below
0.85. The under-coverage problem becomes more severe as $M$ increases because
of the well-known downward bias in kernel variance estimators that reflects
the need to estimate ${\beta}_{0}$ and ${\beta}_{1}$. Coverages of DK and CHS
using $\hat{M}$ are similar to the smaller bandwidth cases, e.g. $M=2$ or
$3$, which makes sense given that the average $\hat{M}$ across
replications is 2.6 (about 0.1 in terms of $\hat{b}$). However, as the
note to the table indicates, large values of $\hat{M}$ can occur in which
case $\hat{b}$ is not close to zero. Because they are bias corrected, the
BCCHS and DKA variance estimators provide coverage that is less sensitive to
the bandwidth. This is particularly true for DKA. If the plug-in fixed-$b$
critical values are used, coverages are closest to 0.95 and very stable across
bandwidths with DKA having the best coverage. Because the CHS variance
estimator is not guaranteed to be positive definite, we report the number of
times that CHS/BCCHS estimates are negative out of the 10,000 replications. In
Table 1 there were no cases where CHS/BCCHS estimates are negative.

\begin{table}[th]
\label{tabledgp1_025} \centering{ \begin{threeparttable}
\begin{tabular}
[c]{cc|cc|cccccc|c}%
\multicolumn{11}{c}{Table 5.2: Sample Coverage Probabilities (\%), Nominal Coverage
95\%}\\
\multicolumn{10}{c}{$N=T=25$; DGP(2): $\omega_{\alpha}=\omega
_{\varepsilon}=0.25$, $\omega_{\gamma}=0.5$, $\rho_{\gamma}= 0.25$; POLS.}\\\hline\hline
&  &  &  &  &  & BC- &  & \multicolumn{2}{c|}{fixed-b c.v.} & Count \\\cline{9-10}
\multicolumn{2}{c|}{No Truncation}&$M$ & $b$ & DK & CHS & CHS & DKA & CHS & DKA & CHS$<0$ \\\hline
EHW& 39.9&$\hat{M}$&$\hat{b}$&85.3&85.9&87.8&89.6&89.1&91.1& 0\\
Ci & 40.7&2  & 0.08                  &86.1&86.6&88.1&90.0&89.5&91.3& 0\\
Ct & 87.1&3  & 0.12                  &84.9&85.5&87.6&89.5&89.5&91.2& 0\\
&     &4  & 0.16                  &83.5&84.1&87.1&89.2&89.6&91.5& 0\\
&     &5  & 0.20                  &82.2&82.8&86.4&88.6&89.7&91.4& 0\\
&     &10 & 0.40                  &75.6&76.2&84.1&86.5&89.7&91.6& 0\\
&     &20 & 0.80                  &64.9&65.5&82.5&85.4&89.4&91.5& 0\\
&     &25 & 1.00                  &60.7&61.3&82.6&85.4&89.2&91.5& 0\\ \hline
\end{tabular}
\begin{tablenotes}
\small
\item Note: $\hat{M}$ ranged from 1 to 12, with an average of 2.5 and a median of 2.
\end{tablenotes}
\end{threeparttable}}\end{table}

\begin{table}[th]
\label{tabledgp1_050} \centering{ \begin{threeparttable}
\begin{tabular}
[c]{cc|cc|cccccc|c}%
\multicolumn{11}{c}{Table 5.3: Sample Coverage Probabilities (\%), Nominal Coverage
95\%}\\
\multicolumn{10}{c}{$N=T=25$; DGP(2): $\omega_{\alpha}=\omega
_{\varepsilon}=0.25$, $\omega_{\gamma}=0.5$, $\rho_{\gamma}= 0.5$; POLS.}\\\hline\hline
&  &  &  &  &  & BC- &  & \multicolumn{2}{c|}{fixed-b c.v.} & Count \\\cline{9-10}
\multicolumn{2}{c|}{No Truncation}&$M$ & $b$ & DK & CHS & CHS & DKA & CHS & DKA & CHS$<0$ \\\hline
EHW& 35.2&$\hat{M}$&$\hat{b}$&81.3&81.9&84.0&86.4&86.0&88.2&0\\
Ci & 37.6&2  & 0.08                  &81.6&82.3&83.8&86.0&85.4&87.5&0\\
Ct & 80.7&3  & 0.12                  &80.7&81.3&83.7&86.1&85.9&88.1&0\\
&     &4  & 0.16                  &79.8&80.1&83.2&85.8&86.3&88.4&0\\
&     &5  & 0.20                  &78.4&78.6&82.8&85.6&86.5&88.8&0\\
&     &10 & 0.40                  &71.5&71.7&80.5&83.5&86.8&89.3&0\\
&     &20 & 0.80                  &60.4&60.6&78.7&82.0&86.2&89.2&0\\
&     &25 & 1.00                  &55.9&56.1&78.5&81.9&86.1&89.2&0\\ \hline
\end{tabular}
\begin{tablenotes}
\small
\item Note: $\hat{M}$ ranged from 1 to 25, with an average of 2.8 and a median of 3.
\end{tablenotes}
\end{threeparttable}}\end{table}

\begin{table}[th]
\label{tabledgp1_075} \centering{ \begin{threeparttable}
\begin{tabular}
[c]{cc|cc|cccccc|c}%
\multicolumn{11}{c}{Table 5.4: Sample Coverage Probabilities (\%), Nominal Coverage
95\%}\\
\multicolumn{10}{c}{$N=T=25$; DGP(2): $\omega_{\alpha}=\omega
_{\varepsilon}=0.25$, $\omega_{\gamma}=0.5$, $\rho_{\gamma}= 0.75$; POLS.}\\\hline\hline
&  &  &  &  &  & BC- &  & \multicolumn{2}{c|}{fixed-b c.v.} & Count \\\cline{9-10}
\multicolumn{2}{c|}{No Truncation}&$M$ & $b$ & DK & CHS & CHS & DKA & CHS & DKA & CHS$<0$ \\\hline
EHW& 28.9&$\hat{M}$&$\hat{b}$&71.4&72.2&76.0&79.0&79.2&82.0& 0\\
Ci & 35.1&2  & 0.08                  &70.9&72.3&74.3&76.9&75.7&78.6& 0\\
Ct & 66.7&3  & 0.12                  &71.4&72.5&75.5&78.4&78.0&80.9& 0\\
&     &4  & 0.16                  &71.0&71.8&75.8&79.0&78.9&82.0& 0\\
&     &5  & 0.20                  &70.0&70.6&75.3&78.7&79.7&82.5& 0\\
&     &10 & 0.40                  &63.8&64.2&72.8&77.1&79.4&83.4& 1\\
&     &20 & 0.80                  &53.2&53.6&71.5&76.3&79.5&83.8&13\\
&     &25 & 1.00                  &48.8&49.1&71.5&76.3&79.3&83.8&13\\ \hline
\end{tabular}
\begin{tablenotes}
\small
\item Note: $\hat{M}$ ranged from 1 to 25, with an average of 3.9 and a median of 4.
\end{tablenotes}
\end{threeparttable}}\end{table}

\begin{table}[th]
\label{tabledgp1_075_nt} \centering{ \begin{threeparttable}
\begin{tabular}
[c]{cc|cc|cccccc|c}%
\multicolumn{11}{c}{Table 5.5: Sample Coverage Probabilities (\%), Nominal Coverage
95\%}\\
\multicolumn{10}{c}{$N=T=50$; DGP(2): $\omega_{\alpha}=\omega
_{\varepsilon}=0.25$, $\omega_{\gamma}=0.5$, $\rho_{\gamma}= 0.75$; POLS.}\\\hline\hline
&  &  &  &  &  & BC- &  & \multicolumn{2}{c|}{fixed-b c.v.} & Count \\\cline{9-10}
\multicolumn{2}{c|}{No Truncation}&$M$ & $b$ & DK & CHS & CHS & DKA & CHS & DKA & CHS$<0$ \\\hline
EHW& 17.3&$\hat{M}$&$\hat{b}$&78.8&79.4&81.5&82.5&84.2&85.2&0\\
Ci & 25.9&2  & 0.08                  &78.5&79.1&80.8&81.8&82.7&83.9&0\\
Ct & 66.0&3  & 0.12                  &78.4&79.2&81.5&82.8&83.9&85.5&0\\
&     &4  & 0.16                  &77.5&78.2&81.7&83.0&85.2&86.5&0\\
&     &5  & 0.20                  &76.2&76.9&81.0&82.5&86.0&87.1&0\\
&     &10 & 0.40                  &69.7&70.0&79.2&81.0&86.2&87.8&0\\
&     &20 & 0.80                  &58.5&59.0&77.7&79.4&86.0&87.8&0\\
&     &25 & 1.00                  &54.4&54.7&77.5&79.2&86.0&87.7&0\\ \hline
\end{tabular}
\begin{tablenotes}
\small
\item Note: $\hat{M}$ ranged from 1 to 26, with an average of 5.4 and a median of 5.
\end{tablenotes}
\end{threeparttable}}\end{table}

Tables 5.2 - 5.5 give results for DGP(2) where the latent components enter in
a non-linear way. Tables 5.2 - 5.4 have both sample sizes equal to 25 with
weights across latent components being the same as DGP(1) ($\omega_{\alpha
}=\omega_{\varepsilon}=0.25$, $\omega_{\gamma}=0.5$). Table 5.2 has mild
persistence in $\gamma_{t}$ ($\rho_{\gamma}=0.25$). Table 5.3 has moderate
persistence ($\rho_{\gamma}=0.5$) and Table 5.4 has strong persistence
($\rho_{\gamma}=0.75$). Tables 5.2-5.4 have similar patterns as Table
\ref{tablechs}: confidence intervals with variance estimators non-robust to
individual or time components under-cover with the under-coverage problem
increasing with $\rho_{\gamma}$. With $\rho_{\gamma}=0.25$, CHS has reasonable
coverage (about 0.86) with small bandwidths but under-covers severely with
large bandwidths. BCCHS performs much better because of the bias correction
and fixed-$b$ critical values provide some additional modest improvements. DKA
has better coverage especially when fixed-$b$ critical values are used with
large bandwidths. As $\rho_{\gamma}$ increases, all approaches have increasing
under-coverage problems with DKA continuing to perform best. Table 5.5 has the
same configuration as Table 5.4 but with both sample sizes increased to 50.
Both BCCHS and DKA show some improvements in coverage. This illustrates the
well-known trade-off between the sample size and magnitude of persistence for
accuracy of asymptotic approximations with dependent data. Regarding bandwidth
choice, the data-dependent bandwidth performs reasonably well for CHS, BCCHS,
and DKA. Finally, the chances of CHS/BCCHS being negative are very small but
not zero and chances decrease as both $N$ and $T$ increase.

To show that large values of $\hat{M}$ are not unusual in DGP(2), we
report in Figure 1 the frequency of $\hat{b}$ among the 10,000 Monte Carlo
replications used in Table 5.4. In this case, more than 21\% of replications
have $\hat{b}\geq0.2$. This explains why bias correction and fixed-$b$
critical values noticeably reduce the under-coverage problem when $\widehat
{M}$ is used.

\bigskip

\begin{center}
    \includegraphics[scale=0.125]{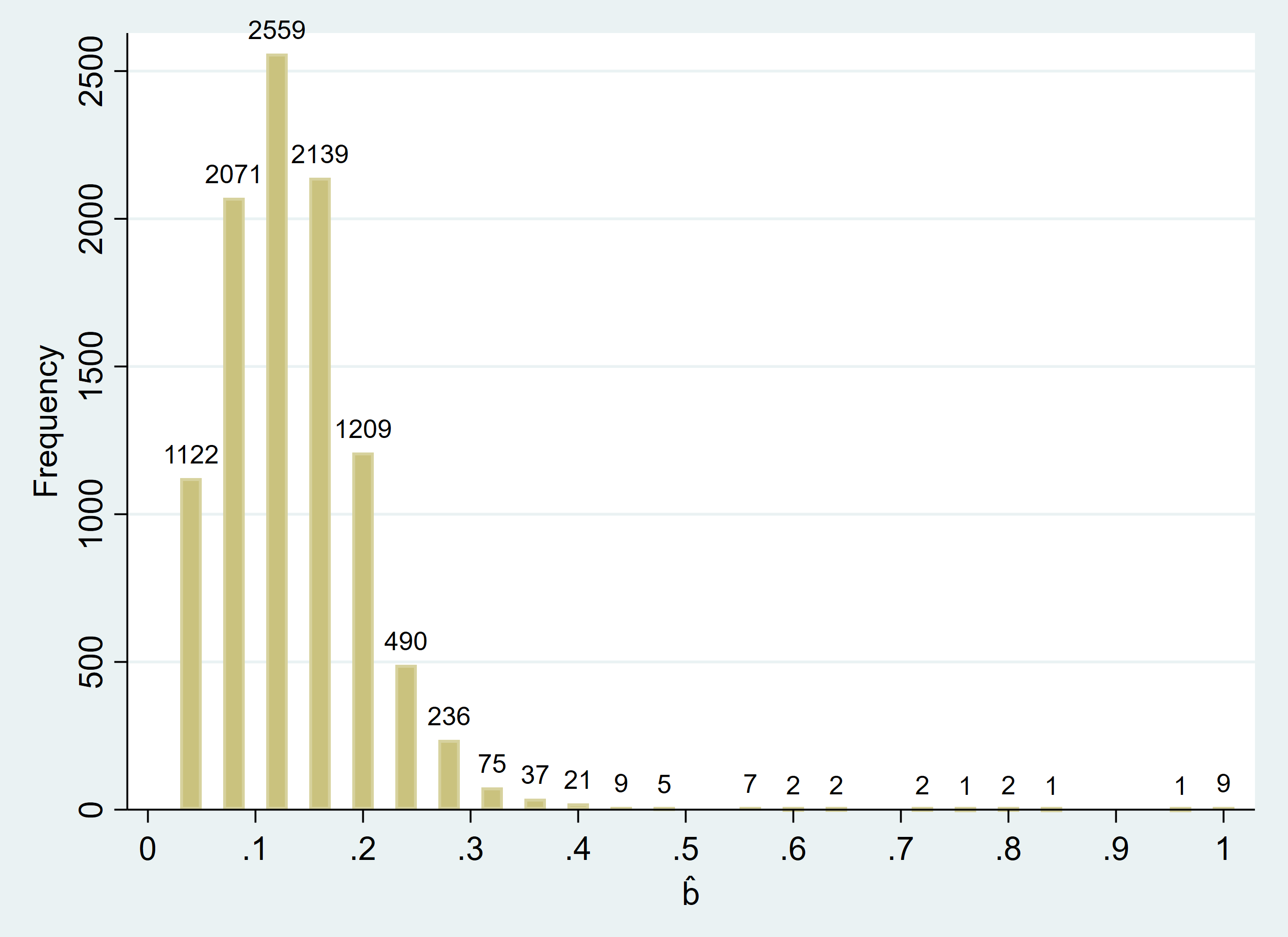} \\
    \textbf{Figure 1}: The Frequency of $\hat{b}$ for Table 5.4.
\end{center}

\bigskip

To show how the relative values of $N$ and $T$ can matter in practice, we
provide additional results for the same DGP as Tables 5.4 and 5.5 for $N$ and
$T$ over a range of values\footnote{We thank a referee for this suggestion,}.
The results are given in Table 5.6. There are two main takeaways from the
table: \textbf{i)} bias correction with and without fixed-$b$ critical values
always improves coverage probabilities relative to the original CHS test,
\textbf{ii)} bias correction alone does slightly better than bias correction
with fixed-$b$ critical values when both $N$, $T$ are extremely small
($N=T=10$).

\begin{table}[th]
\centering{\small \begin{threeparttable}
\begin{tabular}
[c]{cc|ccccccccc|c|c}%
\multicolumn{13}{c}{Table 5.6: Sample Coverage Probabilities (\%), Nominal Coverage 95\%}\\
\multicolumn{13}{c}{$M=\hat{M}$; DGP(2): $\omega_{\alpha}=\omega
_{\varepsilon}=0.25$, $\omega_{\gamma}=0.5$, $\rho_{\gamma}= 0.75$; POLS.}\\\hline\hline
&  &  &  &  &  &  & BC- &  & \multicolumn{2}{c|}{fixed-b c.v.} & Count  &Med  \\\cline{10-11}
$N$ & $T$ & EHW & Ci & Ct & DK & CHS & CHS & DKA & CHS & DKA   & CHS$<0$&$\hat{b}$\\\hline
80 & 160                &12.8&28.8&67.7&86.4&87.1& 88.1&88.7&90.0&90.3&0&0.08\\ %
40 & 160                &19.1&39.7&66.8&85.2&86.7&87.8&88.7&88.8&89.9&0&0.06\\ %
20 & 160                &25.6&48.9&65.7&84.9&87.1&88.2&89.9&88.7&90.4&0&0.05 \\ \hline %
10 & 160                 &33.4&54.5&64.7&83.2&85.9&86.6&89.6&87.3&90.1&0&0.05  \\ \hline %
10 & 80                  &35.0&47.0&64.8&79.3&80.6&82.3&86.3&83.5&87.3&0&0.08 \\ %
10 & 40                  &37.3&42.7&65.7&74.6&74.8&77.2&82.5&78.7&83.7&1&0.13\\ %
10 & 20                  & 43.6&45.2&67.2&69.5&69.1&73.1&80.1&74.5&81.1&15&0.15\\ \hline %
10 & 10                 &53.7&51.9&68.2&63.1&62.8&68.9&79.7& 68.0&77.2&172&0.30 \\ \hline %
20 & 10                 &43.2&45.3&68.0&63.5&64.4&71.5&78.0&73.1&79.4&26&0.30\\ %
40 & 10                 &32.3&35.6&67.6&63.3&64.4&71.0&75.4&75.4&79.3&3&0.30\\ %
80 & 10                &24.5&27.6&68.1&63.8&64.7&71.6&73.9&77.0&79.2&0&0.30\\ \hline %
160 & 10                 &17.9&20.8&68.4&64.0&64.5&71.8&73.0&78.2&79.5&0&0.30\\ \hline %
160 & 20                &13.3&16.8&66.9&70.3&70.7&75.0&75.6&80.1&80.7& 0&0.15\\ %
160 & 40                 &10.3&14.9&66.5&76.9&77.2&79.8&80.4&83.3&83.7&0&0.15\\ %
160 & 80                 &10.7&18.5&66.9&81.5&81.9& 83.9&84.4&86.3&86.8&0&0.09\\\hline
\end{tabular}
\end{threeparttable}}\end{table}

The middle row gives results for both $N$ and $T$ equal to 10. Here the
under-coverage problem is substantial for the original CHS test. Bias
correction helps especially if the DKA estimator is used. Interestingly,
fixed-$b$ critical values help relative to CHS but less than bias correction
alone. This is not surprising because the simulated critical values are
functions of variance estimators based on small sample sizes. Going up the
rows maintains $N=10$ with $T$ increasing to $160$. As expected, coverages
approach 0.95 as $T$ increases. The top four rows have $T=160$ with $N$
increasing from $10$ to $80$. With $T$ fixed and $N$ increasing, the first
three tests that fail to capture within-time/cross-sectional dependence have
deteriorating coverage (undercoverage). In contrast the DK and CHS tests
perform well in those cases; bias correction with fixed-$b$ critical values
continues to provide further improvements.

Going down from the middle rows shows what happens as $N$ increases when $T$
is small ($T=10$). Coverage of the original CHS tests remains quite low as $N$
increases. Bias correction without fixed-$b$ critical values improves coverage
rates but coverage does not improve as $N$ increases. Bias correction with
fixed-$b$ critical values performs best and \textit{improves} as $N$
increases. The results for DKA are interesting. As $N$ increases,
undercoverage becomes more severe when normal critical values are used,
whereas with fixed-$b$ critical values coverge is best and stable across $N$.
The bottom four rows hold $N$ fixed at $160$ and show what happens as $T$
increases from $10$ to $80$. CHS and the bias-corrected versions show better
coverage as $T$ increases. CHS and DKA with fixed-$b$ critical values perform
best in these cases.

\subsection{Simulation Results for the i.i.d.\ Case}

In Theorems 1 - 3, a non-degeneracy assumption on the components is imposed. A
special case that violates this assumption is i.i.d.\ data in both individual
and time dimensions (random sampling). As we showed in Theorem 4, the
fixed-$b$ limit of the test statistics is different in the i.i.d.\ case. By
setting ${\omega}_{\alpha}=0$, ${\omega}_{\gamma}=0$ in DGP(1), we present
coverage probabilities for the i.i.d case in Table 5.7. There are some
important differences between the coverage probabilities in Table 5.7 relative
to previous tables. First, notice that the coverages using EHW, Ci, and Ct are
close to the nominal level as one would expect. The patterns of coverage
probabilities for CHS, BCCHS and DKA are as predicted by Theorem 4 and the
asymptotic calculations given in Table 4.1. Coverages of CHS are close to 0.89
for small bandwidths and under-coverage problems occur with larger bandwidths.
BCCHS is less prone to under-coverage as the bandwidth increases and plug-in
fixed-$b$ critical values help to reduce, but do not eliminate, the
under-coverage problem. In contrast DKA over-covers regardless of the
bandwidth and whether or not fixed-$b$ critical values are used. As $N$,$T$
get larger, we would expect the coverages of CHS/BCCHS to approach 95\% in the
i.i.d. case (assuming a small bandwidth) but not for DKA where over-coverage
would persist.

\begin{table}[th]
\centering{ \begin{threeparttable}
\begin{tabular}
[c]{cc|cc|cccccc|c}%
\multicolumn{11}{c}{Table 5.7: Sample Coverage Probabilities (\%), Nominal Coverage
95\%}\\
\multicolumn{10}{c}{$N=T=25$, i.i.d: DGP (1) with $\omega_{\alpha}=\omega_{\gamma}=0$ and $\omega
_{\varepsilon}=1$; POLS.}\\\hline\hline
&  &  &  &  &  & BC- &  & \multicolumn{2}{c|}{fixed-b c.v.} & Count \\\cline{9-10}
\multicolumn{2}{c|}{No Truncation}&$M$ & $b$ & DK & CHS & CHS & DKA & CHS & DKA & CHS$<0$ \\\hline
EHW& 94.7&$\hat{M}$&$\hat{b}$&91.0& 88.8& 90.7& 98.9& 91.1& 99.0&  73\\
Ci & 93.1&2  & 0.08                  &92.1& 90.2& 91.3& 99.0& 91.9& 99.1&  35\\
Ct & 93.2&3  & 0.12                  &91.0& 88.8& 90.7& 99.0& 91.0& 99.1&  63\\
&     &4  & 0.16                  &89.6& 87.5& 90.0& 98.9& 90.7& 99.0& 117\\
&     &5  & 0.20                  &88.4& 86.1& 89.3& 98.9& 90.2& 99.1& 174\\
&     &10 & 0.40                  &82.0& 80.8& 87.3& 98.7& 88.8& 98.9& 425\\
&     &20 & 0.80                  &70.9& 70.3& 86.1& 98.5& 87.9& 98.9& 639\\
&     &25 & 1.00                  &66.4& 66.0& 85.9& 98.4& 87.7& 98.9& 641\\ \hline
\end{tabular}
\begin{tablenotes}
\small
\item Note: $\hat{M}$ ranged from 1 to 8, with an average of 2.5 and a median of 2.
\end{tablenotes}
\end{threeparttable}}\end{table}

To gauge the extent to which the under-coverage of CHS/BCCHS and over-coverage
of DKA is caused by the mis-match between the plug-in fixed-$b$ critical
values and the i.i.d. fixed-$b$ limits, we report in Table 5.8 simulated
coverage probabilities using fixed-$b$ critical values based on the limits in
Theorem 4. We see that, regardless of the bandwidth, coverages are much closer
to 95\%. Therefore, a significant portion of the size distortions in Table 5.7
is because of the mis-match.

\begin{table}[th]
\centering
\begin{threeparttable}
\begin{tabular}
[c]{c|cccccccc}%
\multicolumn{9}{c}{Table 5.8: Sample Coverage Probabilities (\%)} \\ \hline\hline 
$M$ & $\hat{M}$ & 2 & 3 & 4 & 5 & 10 & 20 & 25 \\
$b$ & $\hat{b}$ & 0.08 & 0.12 & 0.16 & 0.20 & 0.40 & 0.80 & 1.00 \\\hline
CHS & 92.1 & 92.3 & 92.0 & 92.0 & 92.0 & 92.5 & 92.8 & 92.9\\
DKA & 94.1 & 94.0 & 94.0 & 94.1 & 94.0 & 93.9 & 94.0 & 94.0\\ \hline
\end{tabular}
\begin{tablenotes}
\small
\item Nominal coverage probability: 95\%. Sample size: $N=T=25$. DGP: i.i.d: DGP (1) with $\omega_{\alpha}=\omega_{\gamma}=0$ and $\omega
_{\varepsilon}=1$ (same as Table 5.7). Slope estimator: POLS. Fixed-b critical values from Theorem 4 are used for confidence intervals constructed by both CHS and DKA variance estimators.
\end{tablenotes}
\end{threeparttable}
\end{table}

These results raise a practical question: how does a researcher know which
case is being dealt with? In panel data settings, random sampling is almost
never a reasonable assumption in the time dimension and clustering dependence
often exists due to unobserved heterogeneity in both individual and time
dimensions. Some concern may arise as it is common in practice for empirical
researchers to include fixed-effect dummy variables, also as known as two-way
fixed-effect estimator (TWFE, hereafter), to remove at least some of
dependence generated by individual and time unobserved heterogeneity, and in
some cases, such as DGP (1), all of the dependence structure would be removed
and we are back to the i.i.d case. However, it is important to note that DGP
(1) is a very special case. In general, fixed-effect approaches do not
guarantee the resulting scores to be free from clustering dependence. Indeed,
other data generating mechanisms exist where TWFE will not completely remove
the dependence caused by individual and time components in the score as shown
in \cite{CHS_Restat}. We discuss an example and its implications in the next subsection.

One should also note that if we compare absolute size-distortions, DKA-based
tests are not necessarily more concerning than BCCHS-based tests: while in
opposite directions, the magnitudes of size distortions are mostly comparable
and DKA-based tests do a better job as $M$ increases. Moreover, given that the
DKA-based tests tend to be more conservative, a rejection using a DKA-based
test delivers strong evidence against the null hypothesis. For a researcher
that wants to avoid spurious null rejections (relative to the desired
significance level), then DKA-based tests are preferred. On the other hand, if
a rejection is not obtained with BCCHS tests, this is strong evidence that the
null cannot be rejected. Suppose a rejection is obtained with BCCHS but not
with DKA. In this case a researcher had to balance potential over-rejections
from BCCHS with potential lower power of DKA which depend on the extent to
which the researcher thinks two-way clustering is present in the model.

\subsection{Additional Results for TWFE}

A popular alternative to the pooled OLS estimator is the additive TWFE
estimator where individual and time period dummies are included in
(\ref{reg}). It is well known that individual and time dummies will project
out any latent individual or time components that only linearly enter $x_{it}$
and $u_{it}$ individually (as would be the case in DGP(1)) leaving only
variation from the idiosyncratic component $e_{it\text{.}}$. In this case, we
would expect the sample coverages of CHS and DKA to be similar to the i.i.d
case in Table 5.7. However, under the general component structure
representation, the TWFE transformation may not fully remove the individual
and time components if they enter in a nonlinear manner and we would expect
results for CHS and DKA similar to Tables 5.1-5.6.

As an illustration, in Table 5.9 we report results for the TWFE estimator
using the same configuration as Table 5.4 for DGP(2). The sample coverage
probabilities are different from Table 5.4 but are very similar to the results
in Table 5.7 for the i.i.d. case. Therefore, for DGP(2), the TWFE dummy
variables remove the bulk of the variation from the individual and time components.

\begin{table}[th]
\centering{ \begin{threeparttable}
\begin{tabular}
[c]{cc|cc|cccccc|c}%
\multicolumn{11}{c}{Table 5.9: Sample Coverage Probabilities (\%), Nominal Coverage
95\%}\\
\multicolumn{10}{c}{$N=T=25$; DGP(2): $\omega_{\alpha}=\omega
_{\varepsilon}=0.25$, $\omega_{\gamma}=0.5$, $\rho_{\gamma}= 0.75$; TWFE.}\\\hline\hline
&  &  &  &  &  & BC- &  & \multicolumn{2}{c|}{fixed-b c.v.} & Count \\\cline{9-10}
\multicolumn{2}{c|}{No Truncation}&$M$ & $b$ & DK & CHS & CHS & DKA & CHS & DKA & CHS$<0$ \\\hline
EHW& 94.2&$\hat{M}$&$\hat{b}$&91.0&90.2&91.5&98.9&91.6&99.1&  57\\
Ci & 93.3&2  & 0.08                  &92.2&91.0&92.1&99.0&92.3&99.1&  26\\
Ct & 93.0&3  & 0.12                  &90.9&89.7&91.1&99.0&91.7&99.0&  53\\
&     &4  & 0.16                  &89.5&88.4&90.5&98.9&91.1&99.2&  88\\
&     &5  & 0.20                  &88.2&87.4&89.5&98.9&90.4&99.1& 137\\
&     &10 & 0.40                  &81.5&81.7&88.3&98.5&89.5&99.1& 331\\
&     &20 & 0.80                  &70.3&71.3&86.7&98.2&88.7&98.8& 517\\
&     &25 & 1.00                  &66.1&67.2&86.3&98.2&88.4&98.9& 523\\ \hline
\end{tabular}
\begin{tablenotes}
\small
\item Note: $\hat{M}$ ranged from 1 to 8, with an average of 2.5 and a median of 2.
\end{tablenotes}
\end{threeparttable}}\end{table}

In contrast \cite{CHS_Restat} provide an example where the TWFE dummy
variables do not remove the component structure. Consider a third DGP given by%
\begin{align*}
\text{DGP}(3):\ ~x_{it}  &  ={\alpha}_{1i}{\gamma}_{2t}+{\alpha}_{2i}{\gamma}%
_{1t}+{\varepsilon}_{it}^{x},\\
u_{it}  &  ={\alpha}_{1i}{\gamma}_{3t}+{\alpha}_{3i}{\gamma}_{1t}%
+{\varepsilon}_{it}^{u},
\end{align*}
where the latent components $\{{\alpha}_{1i},{\alpha}_{2i},{\alpha}%
_{3i},\gamma_{1t},\gamma_{2t},\gamma_{3t},{\varepsilon}_{it}^{x},{\varepsilon
}_{it}^{u}\}$ are $N(0,1)$ random variables that are independent across $i$
and $t$ and independent with each other. As \cite{CHS_Restat} argue, there is
no endogeneity between $x_{it}$ and $u_{it}$, and it is not difficult to show
that $E(x_{it}|{\alpha}_{i})=E(x_{it}|\gamma_{t})=E(u_{it}|{\alpha}%
_{i})=E(u_{it}|\gamma_{t})=0$. While $x_{it}$ and $u_{it}$ do not have the
component structure, the score, $x_{it}u_{it}$, does because $E(x_{it}%
u_{it}|{\alpha}_{i})={\alpha}_{2i}{\alpha}_{3i}$ and $E(x_{it}u_{it}%
|\gamma_{t})=\gamma_{t2}\gamma_{t3}$. Therefore, the TWFE dummy variables will
not remove the component structure from $x_{it}u_{it}$.

Table 5.10 gives results for DGP(3) for TWFE with $N=T=25$. We see that tests
based on variance estimators that are not robust to two-way cluster dependence
have substantial under-coverage problems. The original CHS does a better job
but tends to under-cover with large bandwidths. BCCHS works better and plug-in
fixed-$b$ critical values provide additional improvements in coverage. DKA
works quite well, with small improvements using plug-in fixed-$b$ critical
values, and coverage probabilities are close to 95\%.

\begin{table}[th]
\centering{ \begin{threeparttable}
\begin{tabular}
[c]{cc|cc|cccccc|c}%
\multicolumn{11}{c}{Table 5.10: Sample Coverage Probabilities (\%), Nominal Coverage
95\%}\\
\multicolumn{10}{c}{$N=T=25$; DGP(4); TWFE.}\\\hline\hline
&  &  &  &  &  & BC- &  & \multicolumn{2}{c|}{fixed-b c.v.} & Count \\\cline{9-10}
\multicolumn{2}{c|}{No Truncation}&$M$ & $b$ & DK & CHS & CHS & DKA & CHS & DKA & CHS$<0$ \\\hline
EHW& 61.5&$\hat{M}$&$\hat{b}$&77.4&89.2& 90.7& 93.9&91.2& 94.2&0\\
Ci & 80.1&2  & 0.08                  &78.6&89.8& 90.9& 93.9&91.1& 94.2&0\\
Ct & 80.0&3  & 0.12                  &77.1&88.8& 90.8& 93.7&91.4& 94.0&0\\
&     &4  & 0.16                  &75.8&87.8& 90.7& 93.6&91.4& 94.3&0\\
&     &5  & 0.20                  &74.5&86.9& 90.4& 93.4&91.2& 94.3&0\\
&     &10 & 0.40                  &67.3&81.8& 89.3& 92.9&91.1& 94.2&0\\
&     &20 & 0.80                  &55.9&71.4& 88.7& 92.3&90.8& 94.2&0\\
&     &25 & 1.00                  &51.7&66.2& 88.6& 92.4&90.8& 94.2&0\\ \hline
\end{tabular}
\begin{tablenotes}
\small
\item Note: $\hat{M}$ ranged from 1 to 25, with an average of 2.6 and a median of 2.
\end{tablenotes}
\end{threeparttable}}\end{table}

The results for BCCHS and DKA in Table 5.10 suggest that fixed-$b$ limits
given by (\ref{WaldBCCHSlimit}) and (\ref{tBCCHSlimit}) for POLS can continue to
hold for tests based on the TWFE estimator of $\beta$. Let $\Ddot{x}_{it}$ and
$\Ddot{u}_{it}$ denote the individual and time dummy demeaned versions of
$x_{it}$ and $u_{it}$ respectively. Suppose that $\Ddot{x}_{it}\Ddot{u}_{it}$,
has the individual and time component structure. Because \cite{CHS_Restat}
show that%
\[
\Ddot{x}_{it}\Ddot{u}_{it}=\Tilde{x}_{it}\Tilde{u}_{it}+o_{p}(1),
\]
where $\Tilde{x}_{it}=x_{it}-\text{E}[x_{it}|\alpha_{i}]-\text{E}%
[x_{it}|\gamma_{t}]+\text{E}[x_{it}]$ and $\Tilde{u}_{it}$ is similarly
defined, equivalent versions Theorem 2, (\ref{WaldBCCHSlimit}),
(\ref{tBCCHSlimit}) and Theorem 3 are easily established for the TWFE
estimator provided the stronger exogeniety assumption, $\text{E}[\Tilde
{x}_{it}u_{it}]=0$, holds\footnote{Strict exogeneity over time, $E(u_{it}%
|x_{i1},x_{i2},...,x_{iT})=0$, is sufficient for $E[\Tilde{x}_{it}u_{it}]=0$
to hold.}.

\section{Empirical Application}

We illustrate how the choice of variance estimator affects $t$-tests and
confidence intervals using an empirical example from \cite{Thompson2011}. We
test the predictive power of market concentration on the profitability of
industries where the market concentration is measured by the
Herfindahl-Hirschman Index (\text{HHI}, hereafter). This example features data where
dependence exists in both cross-section and time dimensions with common shocks
being correlated across time.

Specifically, consider the following linear regression model of profitability
measured by $\text{ROA}_{m,t}$, the ratio of return on total assets for industry $m$
at time $t$:
\[
\text{ROA}_{m,t}=\beta_{0}+\beta_{1}\text{ln}(\text{HHI}_{m,t-1})+\beta_{2}\text{PB}_{m,t-1}+\beta
_{3}\text{DB}_{m.t-1}+\beta_{4}\bar{\text{ROA}}_{t-1}+u_{m,t}%
\]
where $\text{PB}$ is the price-to-book ratio, $\text{DB}$ is the dividend-to-book ratio, and
$\bar{\text{ROA}}$ is the market average $\text{ROA}$ ratio.

\begin{table}[th]
\centering{ \begin{threeparttable}
\begin{tabular}
[c]{ccccccccc}%
\multicolumn{9}{c}{Table 6.1: Industry Profitability, 1972-2021: POLS Estimates
and t-statistics}\\\hline\hline
& POLS & \multicolumn{7}{c}{t-statistics}\\\cline{3-9}%
Regressors & Estimates & EHW & Ci & Ct & DK & CHS & BCCHS & DKA\\\hline
$\text{ln}(\text{HHI}_{m,t-1})$    & 0.0097 & 12.42 & 3.93 & 10.57 & 6.40 & 3.76 & 3.58 & 3.30\\
$\text{Price}/\text{Book}_{m,t-1}$ & -0.0001 & -0.15 & -0.09 & -0.13 & -0.07 & -0.07 & -0.06 & -0.05\\
$\text{DIV}/\text{Book}_{m,t-1}$   & 0.0167 & 6.89 & 3.93 & 3.81 & 2.04 & 1.89 & 1.79 & 1.74\\
Market $\text{ROA}_{t-1}$   & 0.6129 & 32.31 & 14.47 & 12.05 & 12.06 & 10.27 & 9.76 & 8.99\\
Intercept & -0.0564 & -8.94 & -2.76 & -7.52 & -4.69 & -2.67 & -2.53 & -2.35\\\hline
\end{tabular}
\begin{tablenotes}
\small
\item Notes: $R^2=0.117$, $\hat{M}=5$.
\end{tablenotes}
\end{threeparttable}}\end{table}

The data set used to estimate the model is composed of 234 industries in the
US from 1972 to 2021. We obtain the annual level firm data from Compustat and
aggregate it to industry level based on Standard Industry Classification (SIC)
codes. The details of data construction can be found in Section 6 and Appendix
B of \cite{Thompson2011}. \begin{table}[th]
\centering{ \begin{threeparttable}
\begin{tabular}
[c]{ccccccc}%
\multicolumn{7}{c}{Table 6.2: Industry Profitability, 1972-2021: POLS, 95\%
Confidence Intervals}\\\hline\hline
&  &  &  &  & \multicolumn{2}{c}{Fixed-b Critical Values}\\\cline{6-7}%
Regressors & EHW & CHS & BCCHS & DKA & CHS & DKA\\\hline
$\text{ln}(\text{HHI}_{m,t-1})$ & (0.0082, & (0.0046, & (0.0044, & (0.0039, & (0.0043, & (0.0039,\\
& 0.0112) & 0.0147) & 0.0150) & 0.0154) & 0.0149) & 0.0153)\\
$\text{Price}/\text{Book}_{m,t-1}$ & (-0.0017, & (-0.0037, & (-0.0039, & (-0.0044, & (-0.0038, & (-0.0043,\\
& 0.0014) & 0.0035) & 0.0036) & 0.0041) & 0.0037) & 0.0041)\\
$\text{DIV}/\text{Book}_{m,t-1}$ & (0.0119, & (-0.0006, & (-0.0015, & (-0.0022, & (-0.0040, & (-0.0047,\\
& 0.0214) & 0.0340) & 0.0349) & 0.0355) & 0.0353) & 0.0360)\\
Market $\text{ROA}_{t-1}$ & (0.5757, & (0.4959, & (0.4898, & (0.4792, & (0.4844, & (0.4734,\\
& 0.6500) & 0.7299) & 0.7360) & 0.7466) & 0.7352) & 0.7459)\\
Intercept & (-0.0687, & (-0.0978, & (-0.0999, & (-0.1034, & (-0.1000, & (-0.1036,\\
& -0.0440) & -0.0149) & -0.0128) & -0.0093) & -0.0124) & -0.0088)\\\hline
\end{tabular}
\begin{tablenotes}
\small
\item Note: $\hat{M}=5$.
\end{tablenotes}
\end{threeparttable}}\end{table}

In Table 6.1, we present the POLS estimates for the five parameters and
$t$-statistics (with the null $\text{H}_{0}:\beta_{j}=0$ for each $j=1,2,...,5$)
based on the various variance estimators. We use the data dependent bandwidth,
$\hat{M}$, in all relevant cases. We can see the $t$-statistics vary
non-trivially across different variance estimators. The estimated coefficient
of $\text{ln}(\text{HHI}_{m,t-1})$ is significant at a 1\% level based on two-sided t-tests
using any standard errors among comparison, including the DKA standard error.
As is discussed in Section 5.3, a rejection using DKA is strong evidence of
market concentration being powerful in predicting the profitability of
industries. On the other hand, the estimated coefficient of $\text{DIV}/\text{Book}$ is
significant at the 5\% significance level in a two-sided test when EHW,
cluster-by-industry, cluster-by-time, and DK variances are used, while it is
only marginally significant when CHS is used and marginally insignificant when
BCCHS and DKA are used.

In Table 6.2 we present 95\% confidence intervals. For CHS/BCCHS and DKA we
give confidence interval using both normal and plug-in fixed-$b$ critical
values. For the bias corrected variance estimators (BCCHS and DKA) the
differences in confidence intervals between normal and fixed-$b$ critical
values are not large consistent with our simulation results.

\begin{table}[th]
\centering
{ \begin{threeparttable}
\begin{tabular}
[c]{ccccccccc}%
\multicolumn{9}{c}{Table 6.3: Industry Profitability, 1972-2021: TWFE estimates
and t-statistics}\\\hline\hline
& TWFE & \multicolumn{7}{c}{t-statistics}\\\cline{3-9}%
Regressors & Estimates & EHW & Ci & Ct & DK & CHS & BCCHS & DKA\\\hline
$\text{ln}(\text{HHI}_{m,t-1})$    & 0.0050 & 4.27 & 1.84 & 3.54 & 2.04 & 1.55 & 1.46 & 1.33\\
$\text{Price}/\text{Book}_{m,t-1}$ & 0.0015 & 1.73 & 1.41 & 1.53 & 1.00 & 1.03 & 0.97 & 0.78\\
$\text{DIV}/\text{Book}_{m,t-1}$   & 0.0056 & 2.69 & 2.33 & 1.98 & 1.11 & 1.10 & 1.03 & 0.95\\\hline
\end{tabular}
\begin{tablenotes}
\small
\item Notes: $R^2=0.27$, $\hat{M}=6$.
\end{tablenotes}
\end{threeparttable}
}\end{table}\begin{table}[th]
\centering
{ \begin{threeparttable}
\begin{tabular}
[c]{ccccccc}%
\multicolumn{7}{c}{Table 6.4: Industry Profitability, 1972-2021: TWFE, 95\%
Confidence Interval}\\\hline\hline
&  &  &  &  & \multicolumn{2}{c}{fixed-b critical values}\\\cline{6-7}%
Regressors & EHW & CHS & BCCHS & DKA & CHS & DKA\\\hline
$\text{ln}(\text{HHI}_{m,t-1})$ & (0.0027, & (-0.0013, & (-0.0017, & (-0.0024, & (-0.0019, & (-0.0025,\\
& 0.0736) & 0.0114) & 0.0118) & 0.0125) & 0.0126) & 0.0133)\\
$\text{Price}/\text{Book}_{m,t-1}$ & (-0.0020, & (-0.0013, & (-0.0015, & (-0.0022, & (-0.0021, & (-0.0029,\\
& 0.0032) & 0.0043) & 0.0045) & 0.0052) & 0.0045) & 0.0052)\\
$\text{DIV}/\text{Book}_{m,t-1}$ & (0.0015, & (-0.0044, & (-0.0050, & (-0.0059, & (-0.0048, & (-0.0057,\\
& 0.0096) & 0.0155) & 0.0161) & 0.0170) & 0.0166) & 0.0175)\\\hline
\end{tabular}
\begin{tablenotes}
\small
\item Note: $\hat{M}=6$.
\end{tablenotes}
\end{threeparttable}
}\end{table}

In Table 6.3, we include the results for TWFE estimator to see how the
inclusion of firm level and time period dummies matter in practice. The
presence of the dummies results in the intercept and $\bar{\text{ROA}}_{t-1}$
being dropped from the regression. Overall, test statistics based on CHS,
BCCHS, and DKA agree with each other in magnitude and they are much smaller
relative to EHW-based test statistics. As we have seen in Table 5.7, when the
scores are independent in both the cross-section and time dimensions, test
statistics based on those non-twoway robust standard errors tend to be smaller
(higher coverage) on average except for DKA-based tests. Because the
non-twoway test statistics in Table 6.3 are \textit{larger} than the CHS/BCCHS
statistics suggests TWFE does not fully remove two-way dependence and two-way
cluster-robust standard errors are appropriate.

The 95\% confidence intervals for TWFE case are presented in Table 6.4.
Confidence intervals tend to be wider with fixed-$b$ critical values. This is
expected given that fixed-$b$ critical values are larger in magnitude than
standard normal critical values.

\section{Conclusion}

\subsection{Summary}

This paper investigates the fixed-$b$ asymptotic properties of the CHS
variance estimator and tests. An important algebraic observation is that the
CHS variance estimator can be expressed as a linear combination of the cluster
variance estimator, \textquotedblleft HAC of averages" estimator, and
\textquotedblleft average of HACs" estimator. Building upon this observation,
we derive fixed-$b$ asymptotic results for the CHS variance estimator when
both the sample sizes $N$ and $T$ tend to infinity. Our analysis reveals the
presence of an asymptotic bias in the CHS variance estimator which depends on
the ratio of the bandwidth parameter, $M$, to the time sample size, $T$. This
bias is multiplicative and leads to simple feasible bias corrected version of
the CHS variance estimator (BCCHS). We propose a second bias corrected
variance estimator, DKA, by dropping the \textquotedblleft HAC of averages"
that is guaranteed to be positive semi-definite. We show that the fixed-$b$
limiting distribution of tests based on CHS, BCCHS and DKA are not
asymptotically pivotal, and we propose a straightforward plug-in method for
simulating fixed-$b$ asymptotic critical values. Overall,\ we propose four
test statistics that build on the CHS test: BCCHS and DKA tests using
chi-square/standard normal critical values, and BCCHS and DKA tests using
plug-in fixed-$b$ critical values\footnote{CHS tests that use simulated
fixed-$b$ critical values are exactly equivalent to BCCHS tests based on
simulated fixed-$b$ critical values because the fixed-$b$ limits explicitly
capture the bias in the CHS variance estimator.}. Extensive simulations
studies are reported that compare finite sample performance of the proposed
approaches with existing approaches in terms of finite sample null coverage
probabilities. The simple bias-correction approaches provide non-trivial
improvements\ in coverage probabilities and bias-correction with plug-in
fixed-$b$ critical values provide additional improvements except in the i.i.d.
case and when both $N$ and $T$ are very small.

\subsection{Empirical Recommendations}

Our results clearly suggest that the bias corrected variance estimators, BCCHS
and DKA, provide more reliable inference in practice with or without plug-in
fixed-$b$ critical values. While plug-in fixed-$b$ critical values involve
some computation cost in practice, we can generally recommend fixed-$b$
critical values be used in practice given that \textbf{i)} fixed-$b$ critical
values improve finite sample coverage probabilities when large bandwidths are
used, \textbf{ii) }data dependent bandwidths can be large, and \textbf{iii)}
coverage probabilities with or without fixed-$b$ critical values are similar
when bandwidths are small. However, there are important exceptions. When both
the cross-section and time sample sizes are very small, then BCCHS and DKA
based tests using plug-in fixed-$b$ critical values could yield slightly worse
empirical null coverages than using chi-square/standard normal critical values
because the plug-in estimators are noisy. Therefore, the choice between using
fixed-$b$ or chi-square/standard normal critical values for BCCHS and DKA
tests depends on the sample sizes in additional to any relevant computational costs.

The choice between tests based on BCCHS and DKA is nuanced. While DKA ensures
positive definiteness and usually provides tests with better empirical null
coverage probabilities, these benefits do not come without a cost. Although
rare in panel settings, if the scores, $x_{it}u_{it}$, are i.i.d. over both
individual and time dimensions, the DKA estimator has a different fixed-$b$
limiting distribution and tests based on the DKA estimator can be
conservative. In contrast, while the BCCHS estimator also has a different
fixed-$b$ limiting distribution in the i.i.d. case, it has correct asymptotic
coverage probabilities when the bandwidth is small. However, if the bandwidth
is not small, CHS and BCCHS tests under-cover in the i.i.d. case. Therefore,
the practical choice between DKA and BCCHS depends on a researcher's
assessment of the data, the model, and the priority of inference. If the data
is thought to be independent in both dimensions, then one should not consider
cluster-robust variances estimators in the first place. If the data is thought
to have individual and serially correlated time cluster dependence and the
researcher places higher priority on controlling over-rejections while having
a conservative test (with the cost of lower power) should there not be cluster
dependence, the DKA estimator is preferred. BCCHS would be preferred if the
additional under-coverage relative to DKA is viewed as reasonable in order to
have higher power should there not be cluster dependence.

\subsection{Further Discussion}

It is important to acknowledge some limitations of our analysis and to
highlight areas of future research. We found that finite sample coverage
probabilities of all confidence intervals exhibit under-coverage problems when
the autocorrelation of the time effects becomes strong relative to the time
sample size. In such cases, potential improvements resulting from the
fixed-$b$ adjustment is limited. Part of this limitation arises because the
test statistics are not asymptotically pivotal, necessitating plug-in
simulation of critical values. The estimation uncertainty in the plug-in
estimators can introduce sampling errors to the simulated critical values that
can be acute when persistence is strong. Finding a variance estimator that
results in a pivotal fixed-$b$ limit would help address this problem although
appears to be challenging.

An empirically relevant question is whether the component structure is a good
approximation when the component representation in Assumption 1 is not exact.
Ideally, inferential theory should be studied under a DGP where the dependence
is generated not only through individual and time components but also through
the idiosyncratic component. Obtaining fixed-$b$ results for this
generalization appears challenging. Some unreported simulation results point
to some theoretical conjectures but a formal analysis is beyond the scope of
this paper and is left for future research.

A second empirically relevant case we do not address in this paper is the
unbalanced panel data case. There are several challenges in establishing
formal fixed-$b$ asymptotic results for unbalanced panels. Unbalanced panels
have time sample sizes that are potentially different across individuals and
this potentially complicates the choice of bandwidths for the
individual-by-individual variance estimators in the average of HACs component
of the variance. For the Driscoll-Kraay component, the averaging by time would
have potentially different cross-section sample sizes for each period.
Theoretically, obtaining fixed-$b$ results for unbalanced panels also depends
on how the missing data is modeled. For example, one might conjecture that if
missing observations in the panel occur randomly (missing at random), then
extending the fixed-$b$ theory would be straightforward. While that is true in
pure time series settings (see \citealp{rho2019heteroskedasticity}), the presence
of the individual and time random components in the panel setting complicate
things due to the fact that the asymptotic behavior of the components in the
partial sums is very different from the balanced panel case. Obtaining useful
results for the unbalanced panel case is challenging and is a focus of ongoing research.

\section{Acknowledgement}
We thank three anonymous referees, Antonio Galvao, Jeff Wooldridge, Bruce Hansen, Harold Chiang, and participants in AMES 2023 in Beijing and MEG 2022 in East Lansing for helpful comments and suggestions.

\bibliographystyle{elsarticle-harv}
\bibliography{fixed-b_chs}
\renewcommand{\theequation}{A.\arabic{equation}}

\section*{Appendix}





\noindent\textbf{\textit{Proof of Theorem 1}}: Consider $\sqrt{N}(\widehat{\theta}-\theta)$ with the
component structure representation:
\begin{equation}
\sqrt{N}\left(  \widehat{\theta}-\theta\right)  =\frac{1}{\sqrt{N}}\sum
_{i=1}^{N}{a_{i}}+\sqrt{\frac{N}{T}}\frac{1}{\sqrt{T}}\sum_{t=1}^{T}{g_{t}%
}+\frac{1}{\sqrt{N}T}\sum_{i=1}^{N}{\sum_{t=1}^{T}{e_{it}.}} \label{A.1}%
\end{equation}
Under the same set of assumptions, we can apply Theorem 1 of \cite{CHS_Restat}
giving%
\begin{align}
\text{Var}\left(  \frac{1}{\sqrt{N}}\sum_{i=1}^{N}{a_{i}}\right)   &
=\Lambda_{a}\Lambda_{a}^{\prime},\ \ \left\Vert \Lambda_{a}\Lambda_{a}%
^{\prime}\right\Vert <\infty,\label{A.2}\\
\text{Var}\left(  \frac{1}{\sqrt{T}}\sum_{t=1}^{T}{g_{t}}\right)   &
\rightarrow\Lambda_{g}\Lambda_{g}^{\prime},\ \ \left\Vert \Lambda_{g}%
\Lambda_{g}^{\prime}\right\Vert <\infty,\label{A.3}\\
\text{Var}\left(  \frac{1}{\sqrt{NT}}\sum_{i=1}^{N}{\sum_{t=1}^{T}{e_{it}}%
}\right)   &  \rightarrow\Lambda_{e}\Lambda_{e}^{\prime},\ \ \left\Vert
\Lambda_{e}\Lambda_{e}^{\prime}\right\Vert <\infty. \label{A.4}%
\end{align}
Thus, ${a_{i}}=\text{E}[y_{it}-\theta|\alpha_{i}]$ is a sequence of i.i.d
random vectors with zero mean and finite variance $\Lambda_{a}\Lambda
_{a}^{\prime}$. Then, the Lindeberg-L\'{e}vy CLT applies to the first sum in
(\ref{A.1}): as $N\rightarrow\infty$,
\begin{equation}
\frac{1}{\sqrt{N}}\sum_{i=1}^{N}{a_{i}}\overset{d}{\rightarrow}N(0,\Lambda
_{a}\Lambda_{a}^{\prime}). \label{A.5}%
\end{equation}
Consider the second sum in (\ref{A.1}) where $g_{t}=\text{E}[y_{it}%
-\theta|\gamma_{t}]$ is strictly stationary and is an $\alpha$-mixing sequence
with mixing coefficients $\alpha_{g}(\ell)\leq\alpha_{\gamma}(\ell)$ by
Theorem 14.12 of \cite{Hansen2022}, and so $\alpha_{g}(\ell)$ satisfies a
summation condition as follows: for some $s>1$ and $\delta>0$, and for
$K\in(0,\infty)$, there exists integer $N_{K}$ such that
\begin{align*}
\sum_{\ell=1}^{\infty}\alpha_{g}(\ell)^{1-1/2(s+\delta)}  &  \leq\sum_{\ell
=1}^{\infty}\alpha_{\gamma}(\ell)^{1-1/2(s+\delta)}=\sum_{\ell=1}^{N_{K}%
}\alpha_{\gamma}(\ell)^{1-1/2(s+\delta)}+\sum_{\ell=N_{K}+1}^{\infty}\left(
\frac{O(\ell^{-\lambda})}{\ell^{-\lambda}}\ell^{-\lambda}\right)
^{1-1/2(s+\delta)}\\
&  <N_{K}+K\sum_{\ell=1}^{\infty}\left(  \ell^{-\frac{2s}{s-1}}\right)
^{1-1/2(s+\delta)} <\infty.
\end{align*}
Then, by Theorem 16.4 of \cite{Hansen2022} we have as $N,T\rightarrow\infty$,
for $r\in(0,1]$,
\begin{equation}
\frac{1}{\sqrt{T}}\sum_{t=1}^{\left[  rT\right]  }{g_{t}}\Rightarrow
\Lambda_{g}W_{k}\left(  r\right)  , \label{A.6}%
\end{equation}
where $[rT]$ denotes the integer part of $rT$ and $W_{k}\left(  r\right)  $ is
a $k\times1$ vector of standard Wiener process.

As for the third sum, we have $\text{Var}\left(  \frac{1}{\sqrt{N}T}\sum
_{i=1}^{N}{\sum_{t=1}^{T}{e_{it}}}\right)  \rightarrow0$ by (\ref{A.4}). Then,
we can apply Chebyshev's inequality for random variables to show that each
component of the random vector $\frac{1}{\sqrt{N}T}\sum_{i=1}^{N}{\sum
_{t=1}^{T}{e_{it}}}$ converges to $0$ in probability as $N,T\rightarrow\infty$
and so
\begin{equation}
\frac{1}{\sqrt{N}T}\sum_{i=1}^{N}{\sum_{t=1}^{T}{e_{it}}}\overset
{p}{\rightarrow}0. \label{A.7}%
\end{equation}
Combining (\ref{A.5}), (\ref{A.6}), (\ref{A.7}), we have
\[
\sqrt{N}\left(  \widehat{\theta}-\theta\right)  \Rightarrow\Lambda_{a}%
z_{k}+\sqrt{c}\Lambda_{g}W_{k}(1)\ as\ N,T\rightarrow\infty,
\]
and we conclude that $z_{k}$ is independent from $W_{k}(r)$ since $\{a_{i}\}$
and $\{g_{t}\}$ are independent to each other, proving (i) of Theorem 1.

Next, consider (\ref{Shatbar}), scaled by $\frac{1}{\sqrt{N}T}$:
\begin{equation}
\frac{1}{\sqrt{N}T}{\widehat{\bar{S}}}_{[rT]}=\sqrt{\frac{N}{T}}\frac{1}%
{\sqrt{T}}\sum_{t=1}^{\left[  rT\right]  }{\left(  g_{t}-\bar{g}\right)
}+\frac{1}{\sqrt{N}T}\sum_{i=1}^{N}{\sum_{t=1}^{\left[  rT\right]  }{\left(
e_{it}-\bar{e}\right)  .}} \label{A.8}%
\end{equation}
By (\ref{A.7}), we have the second partial sum of (\ref{A.8}) converges to 0
in probability. Combined with the results from (\ref{A.6}) we obtain,
\begin{equation}
\frac{1}{\sqrt{N}T}{\widehat{\bar{S}}}_{[rT]}\Rightarrow\sqrt{c}\Lambda
_{g}\left(  W_{k}(r)-rW_{k}(1)\right)  =\sqrt{c}\Lambda_{g}\widetilde{W}%
_{k}(r), \label{A.9}%
\end{equation}
as $N,T\rightarrow\infty$. Note that for each $t=1,...,T-1$, we can map $t$ to
$[r_{t}T]$ for some $r_{t}\in\left[  \frac{t}{T},\frac{t+1}{T}\right)  $.
Similarly, we can map $t+M$ to $[(r_{t}+b)T]$ where $b=M/T$ and $[r_{t}T]=t$
for $t=1,...,T-M-1$. Using (\ref{A.9}), we have $\frac{1}{\sqrt{N}T}%
{\widehat{\bar{S}}}_{t}\Rightarrow\sqrt{c}\Lambda_{g}\widetilde{W}_{k}(r_{t})$
for each $t=1,...,T-1$ and $\frac{1}{\sqrt{N}T}{\widehat{\bar{S}}}%
_{t+M}\Rightarrow\sqrt{c}\Lambda_{g}\widetilde{W}_{k}(r_{t}+b)$ for each
$t=1,...,T-M-1$. Note that we can take $r_{t}=t/T$, then as $N,T\rightarrow
\infty$, we have
\begin{align*}
\frac{1}{NT^{3}}\sum_{t=1}^{T-1}{{\widehat{\bar{S}}}_{t}}{\widehat{\bar{S}}%
}_{t}^{\prime}=  &  \sum_{r_{t}={1/T}}^{(T-1)/T}{\frac{1}{\sqrt{N}T}%
{\widehat{\bar{S}}}_{[r_{t}T]}}\frac{1}{\sqrt{N}T}{\widehat{\bar{S}}}%
_{[r_{t}T]}^{\prime}\Rightarrow c\Lambda_{g}\int_{0}^{1}\widetilde{W}%
_{k}(r)\widetilde{W}_{k}(r)^{\prime}dr\Lambda_{g},\\
\frac{1}{NT^{3}}\sum_{t=1}^{T-M-1}{{\widehat{\bar{S}}}_{t}}{\widehat{\bar{S}}%
}_{t+M}^{\prime}=  &  \sum_{r_{t}={1/T}}^{(T-M-1)/T}{\frac{1}{\sqrt{N}%
T}{\widehat{\bar{S}}}_{[r_{t}T]}}\frac{1}{\sqrt{N}T}{\widehat{\bar{S}}%
}_{[(r_{t}+b)T]}^{\prime}\Rightarrow c\Lambda_{g}\int_{0}^{1-b}\widetilde
{W}_{k}(r)\widetilde{W}_{k}(r+b)^{\prime}dr\Lambda_{g}.
\end{align*}
Using the results above, we obtain the fixed-$b$ joint limit of (\ref{CHSps2}%
):
\begin{align}
&  \frac{N}{N^{2}T^{2}}\left\{  \frac{2}{M}\sum_{t=1}^{T-1}{{\widehat{\bar{S}%
}}_{t}}{\widehat{\bar{S}}}_{t}^{\prime}-\frac{1}{M}\sum_{t=1}^{T-M-1}{\left(
{\widehat{\bar{S}}}_{t}{\widehat{\bar{S}}}_{t+M}^{\prime}+{\widehat{\bar{S}}%
}_{t+M}{\widehat{\bar{S}}}_{t}^{\prime}\right)  }\right\} \nonumber\\
&  \Rightarrow c\Lambda_{g}\left\{  \frac{2}{b}\int_{0}^{1}{{\widetilde{W}}%
}_{k}{\left(  r\right)  {\widetilde{W}}}_{k}{\left(  r\right)  }{^{\prime}%
dr}-\frac{1}{b}\int_{0}^{1-b}{\left[  {\widetilde{W}}_{k}\left(  r\right)
{\widetilde{W}}_{k}{\left(  r+b\right)  }^{\prime}+{\widetilde{W}}_{k}\left(
r+b\right)  {\widetilde{W}}_{k}{\left(  r\right)  }^{\prime}\right]
}dr\right\}  \Lambda_{g}^{\prime}\nonumber\\
&  =c\Lambda_{g}P\left(  b,{\widetilde{W}}_{k}\left(  r\right)  \right)
\Lambda_{g}^{\prime}. \label{A.10}%
\end{align}
Note that the last term of (\ref{CHSps3}) is canceled out with (\ref{CHSps1}).
The rest of the terms of (\ref{CHSps3}) are functions of the partial sums
defined in (\ref{Shati}). Consider (\ref{Shati}) evaluated at $t=[rT]$ and
scaled by $\frac{1}{T}$:
\[
\frac{1}{T}\widehat{S}_{i,[rT]}=\frac{[rT]}{T}\left(  a_{i}-\bar{a}\right)
+\frac{1}{T}\sum_{t=1}^{[rT]}{\left(  g_{t}-\bar{g}\right)  }+\frac{1}{T}%
\sum_{t=1}^{[rT]}{\left(  e_{it}-\bar{e}\right)  },
\]

We first consider fixed-$N$ and large-$T$ asymptotic results. As
$T\rightarrow\infty$ while fixing $N$, $\frac{[rT]}{T}\left(  a_{i}-\bar
{a}\right)  \overset{p}{\rightarrow}r\left(  a_{i}-\bar{a}\right)  $ and
$\frac{1}{T}\sum_{t=1}^{[rT]}{\left(  g_{t}-\bar{g}\right)  }\overset
{p}{\rightarrow}0$ by (\ref{A.6}). Note that
\[
\text{Var}\left(  \frac{1}{T}\sum_{t=1}^{[rT]}{e_{it}}\right)  =\frac{1}%
{T}\sum_{l=-([rT]-1)}^{[rT]-1}\left(  \frac{[rT]}{T}-\frac{|l|}{T}\right)
\text{E}(e_{it}e_{i,t+l})=\frac{r}{T}\Lambda_{e}\Lambda_{e}^{\prime}(1+o(1)).
\]
By Chebyshev's inequality, we have
\[
\frac{1}{T}\sum_{t=1}^{[rT]}{e_{it}}\overset{p}{\rightarrow}0\text{\ \ as\ \ }%
T\rightarrow\infty.
\]
Therefore, we conclude that
\[
\frac{1}{T}\widehat{S}_{i,[rT]}\overset{p}{\rightarrow}r\left(  a_{i}-\bar
{a}\right)  \text{\ \ as\ \ }T\rightarrow\infty,
\]
which in turn gives that%
\[
\frac{1}{T^{2}}\frac{2}{[bT]}\sum_{t=1}^{T-1}\widehat{S}_{it}\widehat{S}%
_{it}^{\prime}\overset{p}{\rightarrow}\frac{2}{b}\int_{0}^{1}r^{2}\left(
a_{i}-\bar{a}\right)  \left(  a_{i}-\bar{a}\right)  ^{\prime}dr=\frac{2}%
{3b}\left(  a_{i}-\bar{a}\right)  \left(  a_{i}-\bar{a}\right)  ^{\prime
}\text{\ \ as\ \ }T\rightarrow\infty.
\]
So, if we let $T\rightarrow\infty$ and then $N\rightarrow\infty$ sequentially,
we have%
\[
\frac{1}{NT^{2}}\sum_{i=1}^{N}\frac{2}{[bT]}\sum_{t=1}^{T-1}\widehat{S}%
_{it}\widehat{S}_{it}^{\prime}\overset{p}{\rightarrow}\frac{2}{3b}\frac{1}%
{N}\sum_{i=1}^{N}\left(  a_{i}-\bar{a}\right)  \left(  a_{i}-\bar{a}\right)
^{\prime}\text{ \ as \ }T\rightarrow\infty
\]%
\[
=\frac{2}{3b}\frac{1}{N}\sum_{i=1}^{N}\left(  a_{i}a_{i}^{\prime}-a_{i}\bar
{a}^{\prime}-\bar{a}a_{i}^{\prime}+\bar{a}\bar{a}^{\prime}\right)  \overset
{p}{\rightarrow}\frac{2}{3b}\text{E}(a_{i}a_{i}^{\prime})=\frac{2}{3b}%
\Lambda_{a}\Lambda_{a}^{\prime}\text{ \ as \ }N\rightarrow\infty,
\]
where the last convergence follows from the WLLN. What we obtain here is the
sequential limit of the first term in (\ref{CHSps3}). However, the sequential
limit is not necessarily equal to the joint limit. \cite{Phillips1999} provide
a framework to obtain joint convergence results through sequential convergence
results under certain conditions. Following their approach, we first define
sequential convergence and joint convergence for random matrices defined
below and then introduce a lemma which gives a sufficient condition for
sequential convergence to imply joint convergence.

\begin{definition}
Let $G_{NT}$ be defined as%
\[
G_{NT}:=\frac{1}{N}\sum_{i=1}^{N}G_{iT},
\]
where
\[
G_{iT}:=\frac{1}{T^{2}}\frac{2}{[bT]}\sum_{t=1}^{T-1}\widehat{S}_{it}%
\widehat{S}_{it}^{\prime} \overset{p}{\rightarrow}\frac{2}{3b}\left(
a_{i}-\bar{a}\right)  \left(  a_{i}-\bar{a}\right)  ^{\prime}%
=:G_{i}, \ as \ T\to\infty.
\]
Further, define%
\[
G_{N}:=\frac{1}{N}\sum_{i=1}^{N}G_{i} \overset{p}{\rightarrow}\frac{2}%
{3b}\Lambda_{a}\Lambda_{a}=:G, \ as \ N\to\infty.
\]

\end{definition}

\begin{definition}
(a) A sequence of $k\times k$ matrices ${G_{NT}}$ on $(\Omega,\mathcal{F},P)$
is said to converge in probability sequentially to $G$, if
\[
\lim_{N\rightarrow\infty}\lim_{T\rightarrow\infty}P\left(  \left\Vert
G_{NT}-G\right\Vert >\varepsilon\right)  =0\ \ \forall\varepsilon>0.
\]
(b) Suppose that the $k\times k$ random matrices ${G_{NT}}$ and $G$ are
defined on a probability space $(\Omega,\mathcal{F},P)$. $G_{NT}$ is said to
converge in probability jointly to $G$, if
\[
\lim_{N,T\rightarrow\infty}P\left(  \left\Vert G_{NT}-G\right\Vert
>\varepsilon\right)  =0\ \ \forall\varepsilon>0.
\]

\end{definition}

\begin{lemma}
Suppose there exist random matrices $G_{N}$ and $G$ on the same probability
space as $G_{NT}$ satisfying that, for all $N$, $G_{NT}\overset{p}%
{\rightarrow}G_{N}$ as $T\rightarrow\infty$ and $G_{N}\overset{p}{\rightarrow
}G$ as $N\rightarrow\infty$. Then, $G_{NT}\overset{p}{\rightarrow}G$ jointly
if
\begin{equation}
\underset{N,T}{\lim\sup}P\left(  \left\Vert G_{NT}-G_{N}\right\Vert
>\varepsilon\right)  =0\ \ \forall\varepsilon>0. \label{A.11}%
\end{equation}

\end{lemma}

\noindent Lemma 1 can be proved the same way Lemma 6 of \cite{Phillips1999} is
proved, with the only difference that the vector norm is replaced by a matrix
norm, so the proof is omitted here.

Now we verify condition (\ref{A.11}). By Markov's inequality, Minkowski
inequality (for infinite sum), and the fact that there is no heterogeneity of
$G_{iT}$ and $G_{i}$ across $i$, we have
\begin{align*}
\underset{N,T}{\lim\sup}P\left(  \left\Vert G_{NT}-G_{N}\right\Vert
>\varepsilon\right)  \leq\underset{N,T}{\lim\sup}\frac{1}{\varepsilon}%
\text{E}\left\Vert \frac{1}{N}\sum_{i=1}^{N}G_{iT}-\frac{1}{N}\sum_{i=1}%
^{N}G_{i}\right\Vert \leq\underset{N,T}{\lim\sup}\frac{1}{\varepsilon}%
\text{E}\left\Vert G_{iT}-G_{i}\right\Vert .
\end{align*}
Because $G_{iT}$ converges to $G_{i}$ in probability as $T\rightarrow\infty$,
it suffices to show that for each $i$, $\{G_{iT}\}_{T=1}^{\infty}$ is
uniformly integrable for the last term to converge to 0. Let $\zeta>0$ and
consider $\text{E}\left\Vert G_{iT}\right\Vert ^{1+\zeta}$:%
\begin{align*}
\text{E}\left\Vert G_{iT}\right\Vert ^{1+\zeta}  &  =\text{E}\left\Vert
\frac{1}{T^{2}}\frac{2}{[bT]}\sum_{t=1}^{T-1}\widehat{S}_{it}\widehat{S}%
_{it}^{\prime}\right\Vert ^{1+\zeta}\leq\left(  \frac{1}{T^{2}}\frac{2}%
{[bT]}\sum_{t=1}^{T-1}\left(  \text{E}\left\Vert \widehat{S}_{it}\widehat
{S}_{it}^{\prime}\right\Vert ^{1+\zeta}\right)  ^{\frac{1}{1+\zeta}}\right)
^{1+\zeta}\\
&  \leq\left(  \frac{1}{T^{2}}\frac{2}{[bT]}\sum_{t=1}^{T-1}\left(
\text{E}\left\Vert \widehat{S}_{it}\right\Vert ^{2\left(  1+\zeta\right)
}\right)  ^{\frac{1}{2\left(  1+\zeta\right)  }}\left(  \text{E}\left\Vert
\widehat{S}_{it}\right\Vert ^{2\left(  1+\zeta\right)  }\right)  ^{\frac
{1}{2\left(  1+\zeta\right)  }}\right)  ^{1+\zeta},
\end{align*}
where the first and second inequalities follows from Minkowski's and
H\"{o}lder's inequalities respectively. Let $\zeta=\frac{\delta}{4}$ with
$\delta>0$ from Assumption 1 and consider $\left(  \text{E}\left\Vert \frac
{1}{T}\widehat{S}_{it}\right\Vert ^{2\left(  1+\zeta\right)  }\right)
^{\frac{1}{2\left(  1+\zeta\right)  }}$:
\begin{align*}
&  \left(  \text{E}\left\Vert \frac{1}{T}\widehat{S}_{i,t=[rT]}\right\Vert
^{2(1+\zeta)}\right)  ^{\frac{1}{2(1+\zeta)}} \leq\frac{1}{T}\sum
\limits_{j=1}^{[rT]}(\text{E}\Vert y_{ij}\Vert^{2(1+\zeta)})^{\frac
{1}{2(1+\zeta)}}+\frac{[rT]}{T} \frac{1}{T}\left(  \text{E}\Vert y_{it}%
\Vert^{2(1+\zeta)}\right)  ^{\frac{1}{2(1+\zeta)}}<\infty,
\end{align*}
by Minkowski's inequality and Assumption 2(i). Now we conclude that
$\text{E}\Vert G_{it}\Vert^{1+\zeta}<\infty$, i.e. $G_{it}$ is uniformly
integrable by Theorem 6.13 of \cite{Hansen2022}. By uniform integrability of
$G_{iT}$ and convergence in probability of $G_{iT}$ to $G_{i}$, we have
$L^{1}$ convergence: $\underset{N,T}{\lim\sup}\text{E}\left\Vert G_{iT}%
-G_{i}\right\Vert =0$. Then, condition (\ref{A.11}) follows and we obtain the
joint limit of $\frac{1}{N}\sum_{i=1}^{N}G_{iT}$ as $N,T\rightarrow\infty$.
Specifically, we have%
\[
\frac{1}{N}\sum_{i=1}^{N}G_{iT}=\frac{1}{NT^{2}}\sum\limits_{i=1}^{N}{\frac
{2}{\left[  bT\right]  }\sum\limits_{t=1}^{T-1}{{\widehat{S}}_{it}{\widehat
{S}}_{it}^{\prime}}}=\frac{2}{3b}\Lambda_{a}\Lambda_{a}^{\prime}+o_{p}(1),
\]
as $N,T\rightarrow\infty$. Following similar steps, we obtain joint limits for
the rest of the terms in (\ref{CHSps3}):
\begin{align}
\frac{1}{NT^{2}}\frac{1}{M}\sum\limits_{i=1}^{N}{\sum\limits_{t=1}%
^{T-M-1}{\left(  {\widehat{S}}_{it}\ {\widehat{S}}_{i,t+M}^{\prime}%
+{\widehat{S}}_{i,t+M}{\widehat{S}}_{i,t}^{\prime}\right)  }}  &  =\left(
\frac{2}{3b}+\frac{1}{3}\right)  (1-b)^{2}\Lambda_{a}\Lambda_{a}^{\prime
}+o_{p}(1),\label{A.12}\\
\frac{1}{NT^{2}}\frac{1}{M}\sum_{i=1}^{N}{\sum_{t=T-M}^{T-1}{\left(
{\widehat{S}}_{it}\ {\widehat{S}}_{iT}^{\prime}+{\widehat{S}}_{iT}{\widehat
{S}}_{it}^{\prime}\right)  }}  &  =(2-b)\Lambda_{a}\Lambda_{a}^{\prime}%
+o_{p}(1). \label{A.13}%
\end{align}
Combining the partial-sum representation in (\ref{CHSps1}), (\ref{CHSps2}),
(\ref{CHSps3}) and the results above, we obtain (\ref{omhatCHSlimit}). \qedsymbol{}\\

\noindent\textbf{\textit{Proof of Theorem 2}}: First, rewrite $\sqrt{N}(\widehat{\beta}-\beta)$ using the
component structure representation:%
\begin{align*}
\sqrt{N}(\widehat{\beta}-\beta)  &  =\widehat{Q}^{-1}\left(  \frac{\sqrt{N}%
}{NT}\sum_{i=1}^{N}{\sum_{t=1}^{T}{\left(  a_{i}+g_{t}+e_{it}\right)  }%
}\right) \nonumber\\
&  =\widehat{Q}^{-1}\left[  \frac{1}{\sqrt{N}}\sum_{i=1}^{N}{a_{i}}%
+\sqrt{\frac{N}{T}}\frac{1}{\sqrt{T}}\sum_{t=1}^{T}{g_{t}}+\frac{1}{\sqrt{T}%
}\frac{1}{\sqrt{NT}}\sum_{i=1}^{N}{\sum_{t=1}^{T}{e_{it}}}\right]  .
\end{align*}
Next, by Assumption 3(ii) and H\"{o}lder's inequality we have, for some $s>1$
and $\delta>0$,%
\begin{align*}
\text{E}\left(  \lVert x_{it}u_{it}\rVert^{4(s+\delta)}\right)   &
\leq\text{E}\left(  \lVert x_{it}\rVert^{8(s+\delta)}\right)  ^{1/2}%
\text{E}\left(  \lVert u_{it}\rVert^{8(s+\delta)}\right)  ^{1/2}<\infty,\\
\text{E}\left(  \lVert x_{it}x_{it}^{\prime}\rVert^{4(s+\delta)}\right)   &
\leq\text{E}\left(  \lVert x_{it}\rVert^{8(s+\delta)}\right)  ^{1/2}%
\text{E}\left(  \lVert x_{it}\rVert^{8(s+\delta)}\right)  ^{1/2}<\infty,
\end{align*}
Now, we are back to the case of Assumption 1 (ii). Then, by similar steps as
in the proof of Theorem 1, we have $\left\Vert \Lambda_{a}\Lambda_{a}^{\prime
}\right\Vert <\infty$, $\left\Vert \Lambda_{g}\Lambda_{g}^{\prime}\right\Vert
<\infty$, $\left\Vert \Lambda_{e}\Lambda_{e}^{\prime}\right\Vert <\infty$,
and
\begin{align}
\frac{1}{\sqrt{N}} \sum_{i=1}^{N} a_{i}  &  \overset{d}{\rightarrow} N(0,\Lambda
_{a}\Lambda_{a}^{\prime}),\label{limita}\\
\frac{1}{\sqrt{T}} \sum_{t=1}^{[rT]} g_{t}  &  \Rightarrow\Lambda_{g}
W_{k}(r),\label{limitb}\\
\frac{1}{\sqrt{N}T} \sum_{i=1}^{N} \sum_{t=1}^{T} e_{it}  &  \overset{p}{\rightarrow}
0, \label{limitc}%
\end{align}
as $N,T\to\infty$.

As for $\widehat{Q}$, we can vectorize it and then decompose it in the same
manner as the multivariate mean case:
\begin{align*}
vec(x_{it}x_{it}^{\prime})-vec(Q)  &  =a_{i}^{x}+g_{t}^{x}+e_{it}^{x},\\
vec(\widehat{Q})-vec(Q)  &  =\frac{1}{N}\sum_{i=1}^{N}a_{i}^{x}+\frac{1}%
{T}\sum_{t=1}^{T}g_{t}^{x}+\frac{1}{NT}\sum_{i=1}^{N}\sum_{t=1}^{T}e_{it},
\end{align*}
where $a_{i}^{x}=\text{E}[vec(x_{it}x_{it}^{\prime})-vec(Q)|\alpha_{i}]$,
$g_{t}^{x}=\text{E}[vec(x_{it}x_{it}^{\prime})-vec(Q)|\gamma_{t}]$, and
$e_{it}^{x}=vec(x_{it}x_{it}^{\prime})-vec(Q)-a_{i}^{x}-g_{t}^{x}$. Then, we
can apply the results of (\ref{A.1}) - (\ref{A.4}) and the fact that the sums
in (\ref{A.1}) are mutually uncorrelated to conclude that $\text{Var}%
(vec(\widehat{Q}))\rightarrow0$. Then, by Chebyshev's inequality we obtain
$vec(\widehat{Q})\overset{p}{\rightarrow}vec(Q)$, i.e. as $N,T\rightarrow
\infty$,
\begin{equation}
\widehat{Q}\overset{p}{\rightarrow}Q. \label{A.18}%
\end{equation}
Therefore, as $N,T\to\infty$, we have
\[
\sqrt{N}\left(  \widehat{\beta}-\beta\right)  \Rightarrow Q^{-1}\left[
\Lambda_{a}z+\sqrt{c}\Lambda_{g}W(1)\right]
\]
as claimed for the first part of Theorem 2. Next, for the second part we
define the partial sums in the same fashion as (\ref{Shati}) and
(\ref{Shatbar}):
\begin{align*}
{\widehat{S}}_{i,[rT]}  &  =\sum_{t=1}^{[rT]}{x_{it}{\hat{v}}_{it}%
}, \\
{\widehat{\bar{S}}}_{[rT]}  &  = \sum_{i=1}^{N}\sum_{t=1}^{[rT]}%
{x_{it}{\hat{v}}_{it}}. 
\end{align*}

With similar steps as in proving (\ref{A.18}), we can show
\[
\frac{1}{NT}\sum_{i=1}^{N}\sum_{t=1}^{[rT]}x_{it}x_{it}^{\prime} =
\frac{N[rT]}{NT} \frac{1}{N[rT]}\sum_{i=1}^{N}\sum_{t=1}^{[rT]}x_{it}%
x_{it}^{\prime} \overset{p}{\rightarrow}rQ.
\]
and
\[
\frac{1}{T}\sum_{t=1}^{[rT]}x_{it}x_{it}^{\prime}\overset{p}{\rightarrow}rQ.
\]
Therefore, as $N,T\to\infty$, we have
\begin{align*}
\frac{1}{\sqrt{N}T}{\widehat{\bar{S}}}_{[rT]}  &  \Rightarrow r\Lambda
_{a}W_{k}(1)+\sqrt{c}\Lambda_{g}W_{k}(r)-rQ\left(  Q^{-1}\left[  \Lambda
_{a}W_{k}(1)+\sqrt{c}\Lambda_{g}W_{k}(1)\right]  \right) \nonumber\\
&  =\sqrt{c}\Lambda_{g}\widetilde{W}_{k}(r). 
\end{align*}
Then, similarly as it is shown in (\ref{A.10}) we have
\begin{align}
&  \frac{N}{N^{2}T^{2}}\left\{  \frac{2}{M}\sum_{t=1}^{T-1}{{\widehat{\bar{S}%
}}_{t}}{\widehat{\bar{S}}}_{t}^{\prime}-\frac{1}{M}\sum_{t=1}^{T-M-1}{\left(
{\widehat{\bar{S}}}_{t}{\widehat{\bar{S}}}_{t+M}^{\prime}+{\widehat{\bar{S}}%
}_{t+M}{\widehat{\bar{S}}}_{t}^{\prime}\right)  }\right\} \nonumber\\
&  \Rightarrow c\Lambda_{g}\left\{  \frac{2}{b}\int_{0}^{1}{{\widetilde{W}}%
}_{k}{\left(  r\right)  {\widetilde{W}}}_{k}{\left(  r\right)  }{^{\prime}%
dr}-\frac{1}{b}\int_{0}^{1-b}{\left[  {\widetilde{W}}_{k}\left(  r\right)
{\widetilde{W}}_{k}{\left(  r+b\right)  }^{\prime}+{\widetilde{W}}_{k}\left(
r+b\right)  {\widetilde{W}}_{k}{\left(  r\right)  }^{\prime}\right]
}dr\right\}  \Lambda_{g}^{\prime}\nonumber\\
&  =c\Lambda_{g}P\left(  b,{\widetilde{W}}_{k}\left(  r\right)  \right)
\Lambda_{g}^{\prime}. \label{DKlimit}%
\end{align}

For the rest of the terms of $\widehat{\Omega}_\text{CHS}$, we again apply Lemma 1
to obtain the joint limit through the sequential limit. Consider $\frac{1}%
{T}{\widehat{S}}_{i,[rT]}$ using the component representation:
\[
\frac{1}{T}{\widehat{S}}_{i,[rT]}=\frac{[rT]}{T}a_{i}+\frac{1}{T}\sum
_{t=1}^{[rT]}g_{t}+\frac{1}{T}\sum_{t=1}^{[rT]}e_{it}-\frac{1}{T}\sum
_{t=1}^{[rT]}{x_{it}x_{it}^{\prime}\left(  \widehat{\beta}-\beta\right)  }.
\]

Note that $\frac{1}{T}\sum_{t=1}^{[rT]}{e_{it}}\overset{p}{\rightarrow}0$ as
$T\to\infty$ due to $\text{Var}(\frac{1}{T}\sum_{t=1}^{[rT]}{e_{it}})=O(1/T)$
and Chebyshev's inequality. Then, given fixed $N$ and as $T\to\infty$, we
have
\begin{align*}
\widehat{\beta}-\beta &  =\widehat{Q}^{-1}\left[  \frac{1}{N}\sum_{i=1}%
^{N}{a_{i}}+\frac{1}{T}\sum_{t=1}^{T}{g_{t}}+\frac{1}{NT}\sum_{i=1}^{N}%
{\sum_{t=1}^{T}{e_{it}}}\right] \\
&  \overset{p}{\rightarrow}Q^{-1}\bar{a_{i}}%
\end{align*}
and so $\frac{1}{T}{\widehat{S}}_{i,[rT]}\overset{p}{\rightarrow} r \left(  a_{i} -
\bar{a_{i}}\right)  $, which is the same as the sample mean estimator case.
Define
\begin{align*}
G_{NT}:=  &  \frac{1}{N}\sum_{i=1}^{N}G_{iT}, \ \ G_{iT}:=  \frac{1}{T^{2}}\frac{2}{[bT]}\sum_{t=1}^{T-1}\widehat{S}%
_{it}\widehat{S}_{it}^{\prime},\\
G_{i}:=  &  \frac{2}{3b}\left(  a_{i}-\bar{a}\right)  \left(
a_{i}-\bar{a}\right)  ^{\prime}, \ \ G_{N}:=   \frac{1}{N}\sum_{i=1}^{N}G_{i}, \ \ G:=   \frac{2}{3b}\Lambda_{a}\Lambda_{a}%
\end{align*}
where $\widehat{S}_{it} = {\widehat{S}}_{i,[rT]} =\sum_{t=1}^{[rT]}
x_{it}{\hat{v}}_{it}$.

From the Proof of Theorem 1, we know that to prove condition (\ref{A.11}) it
suffices to show the uniform integrability of $\{G_{iT}\}$ for any $i$. For
some $\zeta>0$, we have
\begin{align*}
&  \lim_{M\rightarrow\infty}\sup_{T}\text{E}\left(  \left\Vert G_{iT}%
\right\Vert ;\left\Vert G_{iT}\right\Vert >M\right) \\
\leq &  \lim_{M\rightarrow\infty}\sup_{T}\text{E}\left(  \left\Vert
G_{iT}\right\Vert \left(  \frac{\left\Vert G_{iT}\right\Vert }{M}\right)
^{\zeta};\left\Vert G_{iT}\right\Vert >M\right)  \leq\lim_{M\rightarrow\infty
}\sup_{T}\frac{1}{M^{\zeta}}\text{E}\left(  \left\Vert G_{iT}\right\Vert
^{1+\zeta}\right) \\
\leq &  \lim_{M\rightarrow\infty}\sup_{T}\frac{1}{M^{\zeta}}\left(  \frac
{1}{T^{2}}\frac{2}{[bT]}\sum_{t=1}^{T-1}\left(  \text{E}\left\Vert \widehat
{S}_{it}\right\Vert ^{2\left(  1+\zeta\right)  }\right)  ^{\frac{1}{2\left(
1+\zeta\right)  }}\left(  \text{E}\left\Vert \widehat{S}_{it}\right\Vert
^{2\left(  1+\zeta\right)  }\right)  ^{\frac{1}{2\left(  1+\zeta\right)  }%
}\right)  ^{1+\zeta}.
\end{align*}

Now consider $\left(  \text{E}\left\Vert \widehat{S}_{it}\right\Vert
^{2\left(  1+\zeta\right)  }\right)  ^{\frac{1}{2\left(  1+\zeta\right)  }}$.
By Minkowski's inequality, we have%
\begin{align*}
&  \left(  \text{E}\left\Vert \frac{1}{T}\widehat{S}_{i,t=[rT]}\right\Vert
^{2(1+\zeta)}\right)  ^{\frac{1}{2(1+\zeta)}}=\left(  \text{E}\left\Vert
\frac{1}{T}\sum_{t=1}^{[rT]}x_{it}u_{it}-\frac{1}{T}\sum_{t=1}^{[rT]}%
x_{it}x_{it}^{\prime}(\widehat{\beta}-\beta)\right\Vert ^{2(1+\zeta)}\right)
^{\frac{1}{2(1+\zeta)}}\\
&  \leq\left(  \text{E}\left\Vert \frac{1}{T}\sum_{t=1}^{[rT]}x_{it}%
u_{it}\right\Vert ^{2(1+\zeta)}\right)  ^{\frac{1}{2(1+\zeta)}}+\left(
\text{E}\left\Vert \widehat{Q}_{T}\widehat{Q}_{NT}^{-1}\frac{1}{NT}\sum
_{i=1}^{N}\sum_{t=1}^{T}x_{it}u_{it}\right\Vert ^{2(1+\zeta)}\right)
^{\frac{1}{2(1+\zeta)}}%
\end{align*}
where we denote $\widehat{Q}_{T}=\frac{1}{T}\sum_{t=1}^{T}x_{it}x_{it}%
^{\prime}$ and $\widehat{Q}_{NT}=\frac{1}{NT}\sum_{i=1}^{N}\sum_{t=1}%
^{T}x_{it}x_{it}^{\prime}$. The first term is easily bounded uniformly over
$T$ by applying Minkowski's inequality for infinite sums and
H\"{o}lder's inequality under Assumption 3. By applying H\"{o}lder's
inequality to the second term twice, we have
\begin{align}
&  \text{E}\left\Vert \widehat{Q}_{T}\widehat{Q}_{NT}^{-1}\frac{1}{NT}%
\sum_{i=1}^{N}\sum_{t=1}^{T}x_{it}u_{it}\right\Vert ^{2(1+\zeta)}\nonumber\\
\leq &  \left(  \text{E}\left\Vert \widehat{Q}_{T}\right\Vert ^{4(1+\zeta
)}\right)  ^{1/2}\left(  \text{E}\left\Vert \widehat{Q}_{NT}^{-1}\right\Vert
^{4p(1+\zeta)}\right)  ^{1/p}\left(  \text{E}\left\Vert \frac{1}{NT}\sum
_{i=1}^{N}\sum_{t=1}^{T}x_{it}u_{it}\right\Vert ^{4q(1+\zeta)}\right)  ^{1/q}
\label{threeterms}%
\end{align}
where $\frac{1}{p}+\frac{1}{q}=1$ and $p,q\in\lbrack1,\infty]$. The first term
in (\ref{threeterms}) is bounded with a straightforward application of
Minkowski's inequality for infinite sums and H\"{o}lder's inequality under
Assumption 3. Note that we have shown $\widehat{Q}_{NT}\overset{p}%
{\rightarrow}Q$. By Assumption 3(ii), we have $\Vert Q^{-1}\Vert<\infty$. It
follows that $\Vert\widehat{Q}_{NT}^{-1}\Vert<\infty$ and so $\text{E}%
\left\Vert \widehat{Q}_{NT}^{-1}\right\Vert ^{4p(1+\zeta)}<\infty$ given that
$p$ and $\zeta$ are finite. To determine $p$ and $\zeta$, observe that
\begin{align*}
\left(  \text{E}\left\Vert \frac{1}{NT}\sum_{i=1}^{N}{\sum_{t=1}^{T}%
{x_{it}u_{it}}}\right\Vert ^{4q(1+\zeta)}\right)  ^{\frac{1}{4q(1+\zeta)}}  &
\leq\frac{1}{NT}\sum_{i=1}^{N}\sum_{t=1}^{T}\left(  \text{E}\left\Vert
x_{it}u_{it}\right\Vert ^{4q(1+\zeta)}\right)  ^{\frac{1}{4q(1+\zeta)}},\\
&  \leq\frac{1}{NT}\sum_{i=1}^{N}\sum_{t=1}^{T}\left(  \text{E}\left\Vert
x_{it}\right\Vert ^{8q(1+\zeta)}\text{E}\left\Vert u_{it}\right\Vert
^{8q(1+\zeta)}\right)  ^{\frac{1}{8q(1+\zeta)}}%
\end{align*}
where the first inequality follows from Minkowski's inequality for infinite
sums and the second line follows from H\"{o}lder's inequality. Let $q=s$ and
$\zeta=\delta/q=\delta/s$ where $s$ and $\delta$ are from Assumption 3, then
$p=\frac{s}{s-1}$ and it follows that all three terms in (\ref{threeterms})
are bounded uniformly over $T$.

Therefore, we conclude that $\text{E}\Vert G_{it}\Vert^{1+\zeta}<\infty$ and
so $G_{it}$ is uniformly integrable. Then, condition (\ref{A.11}) is
established and we obtain the joint limit of $G_{NT}$ as $N,T\rightarrow
\infty$:%
\[
G_{NT}\overset{p}{\rightarrow}\frac{2}{3b}\Lambda_{a}\Lambda_{a}^{\prime}.
\]
Following the same steps for the rest of terms in (\ref{CHSps3}) leads to the
same results as (\ref{A.12}) and (\ref{A.13}). \qedsymbol{}\\

\noindent\textbf{\textit{Proof of Theorem 3}}: Following the proof of Theorem 2, the $\widehat{\Omega}%
_\text{DK}$ part of the variance estimator can be rewritten as a function of
partial sums where the function is defined in (\ref{CHSps2}) and so the joint
limit follows from (\ref{DKlimit}):
\begin{align*}
N\widehat{\Omega}_\text{DK}  &  =\frac{1}{NT^{2}}\sum\limits_{t=1}^{T}%
{\sum\limits_{s=1}^{T}k\left(  \frac{\left\vert t-s\right\vert }{M}\right)
\left(  \sum\limits_{i=1}^{N}{{\hat{v}}_{it}}\right)  \left(
\sum\limits_{j=1}^{N}{{\hat{v}}_{js}^{\prime}}\right)  }\\
&  =\frac{1}{NT^{2}}\left\{  \frac{2}{M}\sum_{t=1}^{T-1}{{\widehat{\bar{S}}%
}_{t}}{\widehat{\bar{S}}}_{t}^{\prime}-\frac{1}{M}\sum_{t=1}^{T-M-1}{\left(
{\widehat{\bar{S}}}_{t}{\widehat{\bar{S}}}_{t+M}^{\prime}+{\widehat{\bar{S}}%
}_{t+M}{\widehat{\bar{S}}}_{t}^{\prime}\right)  }\right\} \\
&  \Rightarrow c\Lambda_{g} ^{-1}P\left(  b,{\widetilde{W}}_{k}\left(
r\right)  \right)  \Lambda_{g}^{\prime}.
\end{align*}
For the $\widehat{\Omega}_\text{A}$ part of the variance estimator, under
Assumption 3 we can apply Lemma 2 of \cite{CHS_Restat} giving%
\begin{align}
N\widehat{\Omega}_\text{A}=\frac{1}{NT^{2}}\sum\limits_{i=1}^{N}{\left(
\sum\limits_{t=1}^{T}{{\hat{v}}_{it}}\right)  \left(  \sum\limits_{s=1}%
^{T}{{\hat{v}}_{is}^{\prime}}\right)  }\overset{p}{\rightarrow}\Lambda
_{a}\Lambda_{a}^{\prime}. \label{lambdaA}%
\end{align}
\qedsymbol{}\\

\noindent\textbf{\textit{Proof of Theorem 4}}: Define $\Lambda_{xu}\Lambda_{xu}^{\prime}=\text{Var}%
(x_{it}u_{it})$ and $\Lambda_{xx}\Lambda_{xx}^{\prime}=\text{E}(x_{it}%
x_{it}^{\prime})$. By Jensen's inequality, H\"{o}lder's inequality, and
Assumption 4(ii), we have
\begin{align*}
\left\Vert \Lambda_{xu}\Lambda_{xu}^{\prime}\right\Vert  &  =\left\Vert
\text{E}\left[  (x_{it}u_{it})(x_{it}u_{it})^{\prime}\right]  \right\Vert
\leq\text{E}\left\Vert x_{it}u_{it}\right\Vert ^{2}\leq\left(  \text{E}%
\left\Vert x_{it}\right\Vert ^{4}\right)  ^{1/2}\left(  \text{E}\left\Vert
u_{it}\right\Vert ^{4}\right)  ^{1/2}<\infty,\\
\left\Vert \Lambda_{xx}\Lambda_{xx}^{\prime}\right\Vert  &  =\left\Vert
\text{E}(x_{it}x_{it}^{\prime})\right\Vert \leq\text{E}\left\Vert x_{it}%
x_{it}^{\prime}\right\Vert =\text{E}\left\Vert x_{it}\right\Vert ^{2}<\infty.
\end{align*}
Then, by the WLLN, the functional central limit theorem for i.i.d random
vectors, and Slutsky's Theorem, we have

\begin{align}
\sqrt{NT}(\widehat{\beta}-\beta) =  &  {\left(  \frac{1}{NT}\sum_{i=1}%
^{N}{\sum_{t=1}^{T}{x_{it}x_{it}^{\prime}}}\right)  }^{-1}\left(  \frac
{1}{\sqrt{NT}}\sum_{i=1}^{N}{\sum_{t=1}^{T}{x_{it}u_{it}}}\right)  \Rightarrow
Q^{-1}\Lambda_{xu}W_{k}(1),\label{B.1}\\
\frac{1}{\sqrt{NT}}{\widehat{\bar{S}}}_{[rT]} =  &  \frac{1}{\sqrt{NT}}%
\sum_{i=1}^{N}\sum_{t=1}^{[rT]}x_{it}{u_{it}}-\frac{1}{NT}\sum_{i=1}^{N}%
{\sum_{t=1}^{\left[  rT\right]  }{x_{it}x_{it}^{\prime}\sqrt{NT}\left(
\widehat{\beta}-\beta\right)  }} \Rightarrow\Lambda_{xu}\widetilde{W}_{k}(r),
\label{B.2}%
\end{align}
as $N,T\to\infty$. Then, due to the partial sum representation of
$\widehat{\Omega}_\text{DK}$, we have
\[
NT\widehat{\Omega}_\text{DK}\Rightarrow\Lambda_{xu}P\left(  b,\widetilde{W}%
_{k}(r)\right)  \Lambda_{xu}^{\prime},
\]
where $P\left(  b,\widetilde{W}_{k}(r)\right)  $ is defined the same way as
(\ref{omhatCHSlimit}).

The probability limit of (\ref{CHSps1}) scaled by $NT$ follows from Lemma 3 of
\cite{CHS_Restat}:
\begin{equation}
\frac{1}{NT}\sum_{i=1}^{N}\widehat{S}_{iT}\widehat{S}_{iT}^{\prime}\overset
{p}{\rightarrow}\Lambda_{xu}\Lambda_{xu}^{\prime}. \label{B.3}%
\end{equation}

To derive the joint asymptotic limit of (\ref{CHSps3}), we first obtain its
sequential limit and then apply Theorem 1 of \cite{Phillips1999} to show the
joint limit is given by the sequential limit. By the WLLN and functional CLT
for i.i.d.\ random vectors with finite variance, for given $N$ and as
$T\to\infty$, we have
\begin{align*}
\frac{1}{T}\sum_{t=1}^{[rT]}x_{it}x_{it}^{\prime} \overset{p}{\rightarrow}  &
rQ, \ as \ T \to\infty\\
\frac{1}{\sqrt{T}}\sum_{t=1}^{[rT]} x_{it} u_{it}\Rightarrow &  \Lambda
_{xu}W_{i,k}(r),
\end{align*}
where $W_{i,k}(r)$ is a $k\times1$ vector of standard Wiener process for each
$i$. Therefore, for given $N$ and as $T\to\infty$, we have
\begin{align*}
\sqrt{T}(\widehat{\beta}-\beta) =  &  {\left(  \frac{1}{NT}\sum_{i=1}^{N}%
{\sum_{t=1}^{T}{x_{it}x_{it}^{\prime}}}\right)  }^{-1}\left(  \frac{1}%
{N\sqrt{T}}\sum_{i=1}^{N}{\sum_{t=1}^{T}{x_{it}u_{it}}}\right)  \Rightarrow
Q^{-1}\Lambda_{xu}\bar{Z},\\
\frac{1}{\sqrt{T}}\widehat{S}_{i,[rT]} =  &  \frac{1}{\sqrt{T}}\sum
_{t=1}^{[rT]}x_{it}u_{it}-\frac{1}{T}\sum_{t=1}^{[rT]}x_{it}x_{it}^{\prime
}\sqrt{T}\left(  \widehat{\beta}-\beta\right)  \Rightarrow\Lambda_{xu}\left(
{W}_{i,k}(r) - r\bar{Z}\right)  ,
\end{align*}
for each $i$ and $\bar{Z}=\frac{1}{N}\sum_{i=1}^{N}Z_{i}$ where $Z_{i}$ is a
$k\times1$ vector of standard normal random variables.

Because the convergence of a sequence of matrices $A_{n}$ to some matrix
$A_{0}$ holds if and only if $e^{\prime}A_{n}e$ converges to $eA_{0}e$ for any
comfortable constant vector vector $e$, we can assume without loss of
generality that $k=1$. The sequential limit of the first term of
(\ref{CHSps3}), scaled by $NT$, is obtained as follows:
\begin{align*}
Y_{i,T}  &  := \frac{1}{T}\frac{2}{[bT]}\sum_{t=1}^{T-1}\widehat{S}%
_{it}\widehat{S}_{it}\Rightarrow\Lambda_{xu}\frac{2}{b}\int_{0}^{1}\left(
{W}_{i,k}(r) - r\bar{Z}\right)  ^{2}dr\Lambda_{xu}=: Y_{i},\ as \ T\rightarrow
\infty,\\
\frac{1}{N}\sum_{i=1}^{N}Y_{i}  &  =\frac{2}{b}\Lambda_{xu}\int_{0}^{1}\left[
\frac{1}{N}\sum_{i=1}^{N}\left(  {W}_{i,k}(r) - r\bar{Z}\right)  ^{2}\right]
dr\Lambda_{xu}\overset{p}{\rightarrow} \frac{2}{b} \Lambda_{xu} \int_{0}^{1} r
dr \Lambda_{xu} = \frac{1}{b}\Lambda_{xu}\Lambda_{xu},\ as \ N\rightarrow
\infty,
\end{align*}
where the equality in the second line follows from Tonelli Theorem. Noting
that there is no heterogeneity across $i$ due to i.i.d sequences, the
conditions needed for Theorem 1 of \cite{Phillips1999} reduce to the following
conditions: $(i)\ \limsup_{T\rightarrow\infty}\text{E}|Y_{i,T}|<\infty;
(ii)\ \limsup_{T\rightarrow\infty}|EY_{i,T}-EY_{i}|=0; (iii)\ \limsup
_{N,T\rightarrow\infty}\text{E}\left(  |Y_{i,T}|;|Y_{i,T}>N\varepsilon
|\right)  =0\ \forall\varepsilon>0;\ and \ (iv)\ \limsup_{N\rightarrow\infty
}\text{E}\left(  |Y_{i}|;|Y_{i}>N\varepsilon|\right)  =0\ \forall
\varepsilon>0.$

Therefore, it suffices to show uniform integrability of $Y_{i,T}$ and $Y_{i}$.
Uniform integrability of $Y_{i}$ is trivial since it is equivalent to show
$\text{E}|Y_{i}|<\infty$. To show uniform integrability of $Y_{i,T}$, fix
$\varepsilon>0$. We want to show that $\sup_{N,T}\text{E}|Y_{i,T}|<\infty$ and
there exists $\delta$ such that if $P(A)<\delta$ then $\sup_{N,T}%
\text{E}(|Y_{i,T}|;A)<\varepsilon$. By H\"{o}lder's inequalities, we have
\begin{align*}
\text{E}(|Y_{i,T}|;A)=\frac{2}{[bT]}\sum_{t=1}^{T-1}\text{E}\left(  \left\vert
\frac{1}{T}\widehat{S}_{it}\widehat{S}_{it}\right\vert ;A\right)   &
\leq\frac{2}{[bT]}\sum_{t=1}^{T-1}\left(  \text{E}\left(  \frac{1}{T}%
\widehat{S}_{it}^{2};A\right)  \text{E}\left(  \frac{1}{T}\widehat{S}_{it}%
^{2};A\right)  \right) \\
\text{E}|Y_{i,T}|  &  \leq\frac{2}{[bT]}\sum_{t=1}^{T-1}\left(  \text{E}%
\left(  \frac{1}{T}\widehat{S}_{it}^{2}\right)  \text{E}\left(  \frac{1}%
{T}\widehat{S}_{it}^{2}\right)  \right)
\end{align*}
Thus, it is equivalent to show uniform integrability of $\frac{1}{T}%
\widehat{S}_{it}^{2}$. Notice that $\frac{1}{T}\widehat{S}_{it}^{2}=\frac
{1}{T}{S}_{it}^{2}+o_{p}(1)$ under the consistency of $\widehat{\beta}$ and so
by the asymptotic equivalence lemma, we have
\[
\text{E}\left\vert \frac{1}{T}\widehat{S}_{it}^{2}\right\vert =\text{E}%
\left\vert \frac{1}{T}{S}_{it}^{2}\right\vert \ as\ N,T\rightarrow\infty.
\]
Observe that
\[
\text{E}\left\vert \frac{1}{T}{S}_{it}^{2}\right\vert =\frac{1}{T}%
\text{E}\left(  \sum_{t=1}^{T}x_{it}u_{it}\right)  ^{2}=\text{E}(x_{it}%
u_{it}u_{it}x_{it})\ \forall T,
\]
where the second equality follows from that $\{x_{it}u_{it}\}$ are i.i.d
across $t$. Then, under Assumption 4(ii), there exists some constant
$C<\infty$ such that $\text{E}\left\vert \frac{1}{T}{S}_{it}^{2}\right\vert
<C$. Integrating both sides of this inequality over $A$ gives:
\[
CP(A)>\int_{A}\text{E}\left\vert \frac{1}{T}{S}_{it}^{2}\right\vert
dP=\text{E}\left(  \int_{A}\frac{1}{T}{S}_{it}^{2}dP\right)  =\text{E}\left(
\frac{1}{T}{S}_{it}^{2};A\right)  \ \forall T\in\mathbb{N}%
\]
where the second equality follows from that $\frac{1}{T}S_{it}^{2}\geq0$. So,
if we take $\delta=\varepsilon/C$, then $\sup_{N,T}\text{E}\left(  \frac{1}%
{T}{\widehat{S}}_{it}^{2};A\right)  =\sup_{N,T}\text{E}\left(  \frac{1}{T}%
{S}_{it}^{2};A\right)  <\varepsilon$. It follows that $\{\frac{1}{T}%
{\widehat{S}}_{it}^{2}\}$ is uniformly integrable and so $Y_{i,T}$ is
uniformly integrable. Therefore, Theorem 1 of \cite{Phillips1999} applies and
we obtain $Y_{i,T}\overset{p}{\rightarrow}\frac{1}{b}\Lambda_{xu}\Lambda_{xu}$. Similarly,
the joint fixed-$b$ limit of the rest of the terms of (\ref{CHSps3}) are
obtained as follows:
\begin{align*}
\frac{1}{NT}\frac{2}{[bT]}\sum_{i=1}^{N}\sum_{t=1}^{T-M-1}\widehat{S}%
_{it}\widehat{S}_{i,t+M}  &  \overset{p}{\rightarrow} \frac{(1-b)^{2}}{b}\Lambda_{xu}%
\Lambda_{xu}\ as\ N,T\rightarrow\infty,\\
\frac{1}{NT}\frac{2}{[bT]}\sum_{i=1}^{N}\sum_{t=T-M}^{T}\widehat{S}%
_{it}\widehat{S}_{iT}  &  \overset{p}{\rightarrow} \frac{1-(1-b)^{2}}{b}\Lambda_{xu}%
\Lambda_{xu}\ as\ N,T\rightarrow\infty.
\end{align*}

Arranging the joint limits we obtain above we find that $NT(\widehat{\Omega
}_\text{A}-\widehat{\Omega}_\text{NW})=o_{p}(1)$ under Assumption 4, which combined with
(\ref{B.1}) - (\ref{B.3}) delivers the desired result. $\qedsymbol{}$

\end{document}